\documentclass[aps,prc,floatfix,twocolumn,nofootinbib,superscriptaddress]{revtex4-1}

\usepackage{amsmath,bm}
\usepackage{graphicx}
\usepackage{epstopdf}
\usepackage{subfigure}
\usepackage{epsfig}
\usepackage{amsmath,amssymb,amsfonts}
\usepackage{color}
\usepackage[utf8]{inputenc}
\usepackage{verbatim}
\usepackage{array}
\usepackage{hyperref}
\graphicspath{{figures/}} 

\begin{document}

\title{$Z^0$ boson associated b-jet production in high-energy nuclear collisions}

\author{Sa Wang}
\affiliation{Guangdong Provincial Key Laboratory of Nuclear Science, Institute of Quantum Matter, South China Normal University, Guangzhou 510006, China}
\affiliation{Guangdong-Hong Kong Joint Laboratory of Quantum Matter, Southern Nuclear Science Computing Center, South China Normal University, Guangzhou 510006, China}
\affiliation{Key Laboratory of Quark \& Lepton Physics (MOE) and Institute of Particle Physics, Central China Normal University, Wuhan 430079, China}

\author{Wei Dai}
\affiliation{School of Mathematics and Physics, China University of Geosciences, Wuhan 430074, China}

\author{Ben-Wei Zhang}
\email{bwzhang@mail.ccnu.edu.cn}
\affiliation{Key Laboratory of Quark \& Lepton Physics (MOE) and Institute of Particle Physics, Central China Normal University, Wuhan 430079, China}
\affiliation{Guangdong-Hong Kong Joint Laboratory of Quantum Matter, Southern Nuclear Science Computing Center, South China Normal University, Guangzhou 510006, China}

\author{Enke Wang}
\affiliation{Guangdong Provincial Key Laboratory of Nuclear Science, Institute of Quantum Matter, South China Normal University, Guangzhou 510006, China}
\affiliation{Guangdong-Hong Kong Joint Laboratory of Quantum Matter, Southern Nuclear Science Computing Center, South China Normal University, Guangzhou 510006, China}
\affiliation{Key Laboratory of Quark \& Lepton Physics (MOE) and Institute of Particle Physics, Central China Normal University, Wuhan 430079, China}

\date{\today}

\begin{abstract}
The production of vector boson tagged heavy quark jets potentially provides new tools to probe the jet quenching effect.
In this paper, we present the first theoretical study on the angular correlations~($\Delta\phi_{bZ}$), transverse momentum imbalance~($x_{bZ}$), and nuclear modification factor~($I_{AA}$) of $Z^0$ boson tagged b-jets in heavy-ion collisions, which was performed using a Monte Carlo transport model. We find that the medium modification of the $\Delta\phi_{bZ}$ for $Z^0\,+\,$b-jet has a weaker dependence on $\Delta\phi_{bZ}$ than that for $Z^0\,+\,$jet, and the modification patterns are sensitive to the initial jet $p_T$ distribution. Additionally, with the high purity of the quark jet in $Z^0\,+\,$(b-)jet production, we calculate the momentum imbalance $x_{bZ}$ and the nuclear modification factor $I_{AA}$ of $Z^0\,+\,$b-jet in Pb+Pb collisions. We observe a smaller $\Delta \left\langle x_{jZ} \right\rangle$ and larger $I_{AA}$ of $Z^0\,+\,$b-jet in Pb+Pb collisions relative to those of $Z^0\,+\,$jet, which may be an indication of the mass effect of jet quenching and can be tested in future measurements.
\end{abstract}

\maketitle

\section{Introduction}
\label{sec:Introduction}
The formation of quark-gluon plasma (QGP), which was produced in the early stages of the high-energy nucleus-nucleus collisions at the Relativistic Heavy Ion Collider (RHIC) and the Large Hadron Collider (LHC), offers a new possibility to test quantum chromodynamics (QCD) under an extremely hot and dense deconfined state of nuclear matter. The high-$p_T$
partons~(quarks and gluons) produced in the initial hard scattering strongly interact with the QGP and dissipate their energy to the medium, which is referred to as the jet quenching effect~\cite{Wang:1991xy,Gyulassy:2003mc,Qin:2015srf,Vitev:2008rz,Vitev:2009rd}. Consequently, the ``quenched jet'' observables are used to quantify the properties~\cite{Burke:2013yra} of the hot and dense QCD matter by investigating their medium modifications in heavy-ion collisions (HIC) relative to their p+p baselines.

Recently, the associated production of a vector boson ( photon $\gamma$ or electroweak boson such as $Z^0$ and $W^{+/-}$) and jets~($V+$jet) has been extensively studied both theoretically~\citep{ Catani:2002ny,Boughezal:2015ded,Ridder:2015dxa,Boughezal:2015dva,Peierls:1977ci} and experimentally~\citep{Aad:2013ysa,Aad:2013zba,Aaboud:2016sdm,Chatrchyan:2013mwa,Chatrchyan:2013oda,Khachatryan:2014zya,Khachatryan:2014hpa,Sirunyan:2017jix,Aad:2012en} to test the fundamental properties of QCD and improve the constraints on the parton distribution function~(PDF) of a proton. More importantly, because the vector boson would not involve the strong interaction with the medium and gauge the initial energy of the tagged jets, V+jet is recognized as the ideal probe of the properties of QGP~\citep{Neufeld:2010fj,Wang:1996yh,Dai:2012am,Luo:2018pto,Zhang:2018urd,Chen:2018fqu,Kang:2017xnc,Casalderrey-Solana:2015vaa,KunnawalkamElayavalli:2016ttl, Zhang:2018urd,Wang:2013cia,Qin:2012gp,Chang:2019sae,Chen:2017zte,Chatrchyan:2012gt,Sirunyan:2017jic,Sirunyan:2018ncy,Aaboud:2018anc,Yan:2020zrz}.

In particular, new measurements of the associated production of the $Z^0$ boson and b-jet (denoted as $Z^0\,+\,$b-jet) in p+p collisions at the LHC
have been performed by ATLAS and CMS~\cite{Aad:2011jn,Chatrchyan:2012vr,Chatrchyan:2013zja,Chatrchyan:2013zna,Aad:2014dvb,Chatrchyan:2014dha,Khachatryan:2016iob}, since the final state b-jet associated with the $Z^0$ boson is the dominant background of the associated production of Higgs and $Z^0$ bosons ($Z^0+H\rightarrow Z+b\bar{b}$) within the standard model (SM)~\cite{Chatrchyan:2012ww} and can test many physics scenarios beyond the SM that predict new generation mechanism of b quarks and $Z^0$ bosons~\cite{Chatrchyan:2013zja}. It is noted that in HIC, the $Z^0$ boson tagged b-jet (as the initial energy of the b quark is well gauged by the vector boson and thus its energy loss can be directly obtained) is particularly suitable for exploring the quenching of the heavy flavor jet~\cite{Kartvelishvili:1995fr}. The ``dead-cone'' effect~\cite{ALICE:2021aqk} of heavy quarks in QGP may lead to a smaller energy loss compared to light flavors, which is known as the mass effect of jet quenching and has attracted intense investigations ~\cite{Dokshitzer:2001zm,Zhang:2003wk,Armesto:2003jh,Sharma:2009hn,Xing:2019xae,Djordjevic:2014yka}. At the particle level, the latest measurements indicate that the yield of the B meson appears to be less suppressed than that of the D meson in nucleus-nucleus collisions~\cite{Khachatryan:2016ypw,STAR:2021uzu,PHENIX:2022wim}. However, no clear evidence was found at the full-jet level in the previous experimental measurements, such as the comparison of the $R_{AA}$ between the inclusive jet and b-jet~\cite{Chatrchyan:2013exa,Khachatryan:2016jfl} and the comparison of the $p_T$ imbalance ($x_J$) between inclusive dijets and $b\bar{b}$ dijets~\cite{Sirunyan:2018jju}, except for some preliminary indication in the recent measurement implemented by the ATLAS collaboration \cite{ATLAS:2022fgb}. The possible reasons for this can be manifold; e.g., the large contribution of gluon-initiated b-parton processes may thwart the attempts at solving the problem. Previous studies~\cite{Kartvelishvili:1995fr,Neufeld:2010fj} have indicated that the dominant contribution of the $Z^0$ tagged jet is a quark-initiated jet, and the study of $Z^0\,+\,$b-jet in HIC, especially its different medium modifications compared with that of $Z^0\,+\,$jet, will provide a very useful tool to directly address the mass effect between the light-quark jet and massive bottom jet. Nevertheless, thus far, studies on the associated production of the b-jet and $Z^0$ boson in nucleus-nucleus collisions are lacking.

With this in mind, in this work, we present a Monte Carlo transport simulation including elastic (collisional)~\cite{Cao:2013ita} and inelastic (radiative)~\cite{Guo:2000nz,Zhang:2003yn,Zhang:2004qm,Majumder:2009ge} interactions of the energetic parton in the hot/dense QCD medium, while taking the next-to-leading order (NLO) plus parton shower (PS) generated initial hard parton spectrum~\cite{Gleisberg:2008ta} as input, to study the in-medium modification of the vector boson $Z^0$ tagged b-jets. This framework was employed to describe the heavy-flavor jet production of high-energy nuclear collisions in our previous studies~\cite{Dai:2018mhw,Wang:2019xey,Wang:2020bqz,Wang:2020ukj,Wang:2021jgm,Li:2022tcr}. We first present our numerical results for $Z^0\,+\,$jet and compare them with the available experimental data to test the applicability of our model. Then, we calculate the angular correlations of $Z^0\,+\,$b-jet in A+A collisions and demonstrate that the modifications of these correlations are sensitive to the initial b-jet $p_T$ distribution instead of the azimuthal angle. In contrast to the case of $Z^0\,+\,$jet, the requirement of b-tagging excludes the contribution from multiple jets so that the azimuthal angular correlations of $Z^0\,+\,$b-jet show distinct pattern modifications. With the high purity of the light-quark jet in $Z^0\,+\,$jet events~\cite{Kartvelishvili:1995fr,Neufeld:2010fj}, we expect to address the mass dependence of the jet quenching effects between $Z^0\,+\,$jet and $Z^0\,+\,$b-jet.

The remainder of this paper is organized as follows. In Sec.~\ref{sec:ppbaseline} we present the productions of $Z^0\,+\,$b-jet in p+p collisions calculated via the Monte Carlo event generator and comparisons with experimental data. In Sec.~\ref{sec:eloss}, we discuss our treatments of the jet in-medium evolution in A+A collisions. In Sec.~\ref{sec:results}, we present the simulated results and discussions of the azimuthal angular correlation, transverse momentum imbalance, and nuclear modification factor of $Z^0\,+\,$b-jet in HIC. Finally, Sec.~\ref{sec:summary} summarizes the study.

\section{Associated Production of $Z^0$ Boson and b-jet in p+p Collisions}
\label{sec:ppbaseline}

\begin{figure}[!t]
\begin{center}
\vspace*{-0.2in}
\hspace*{-.1in}
 \subfigure[]{\label{fig:process1}
  \epsfig{file=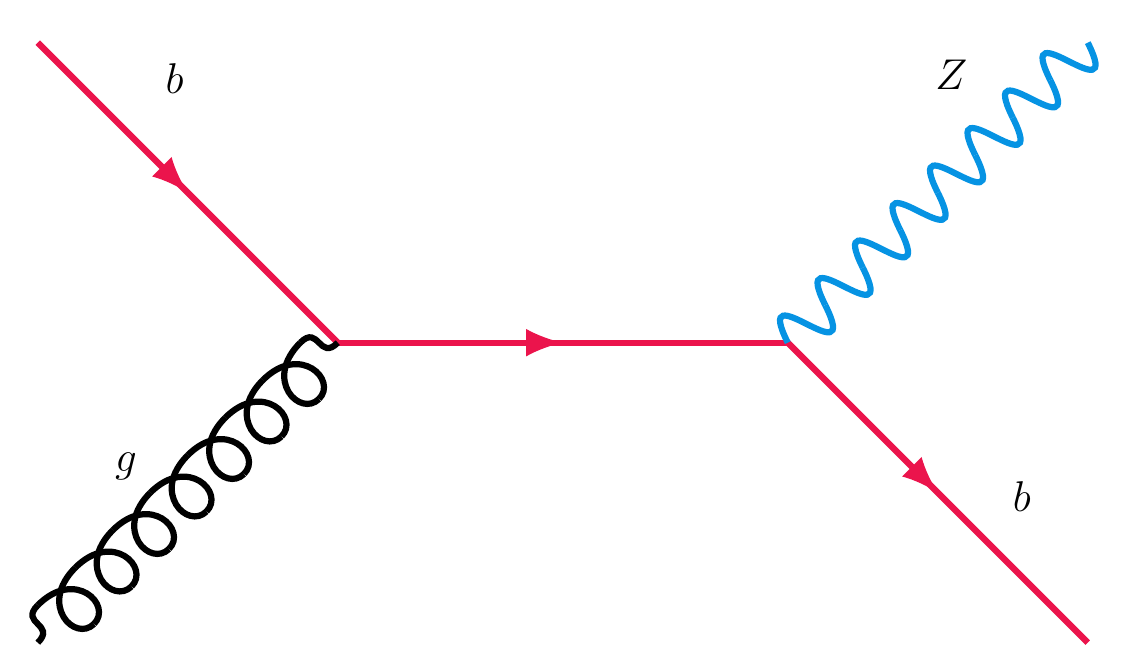, width=0.2\textwidth, clip=}}
  \vspace*{0.2in}
\hspace*{.1in}
 \subfigure[]{\label{fig:process2}
  \epsfig{file=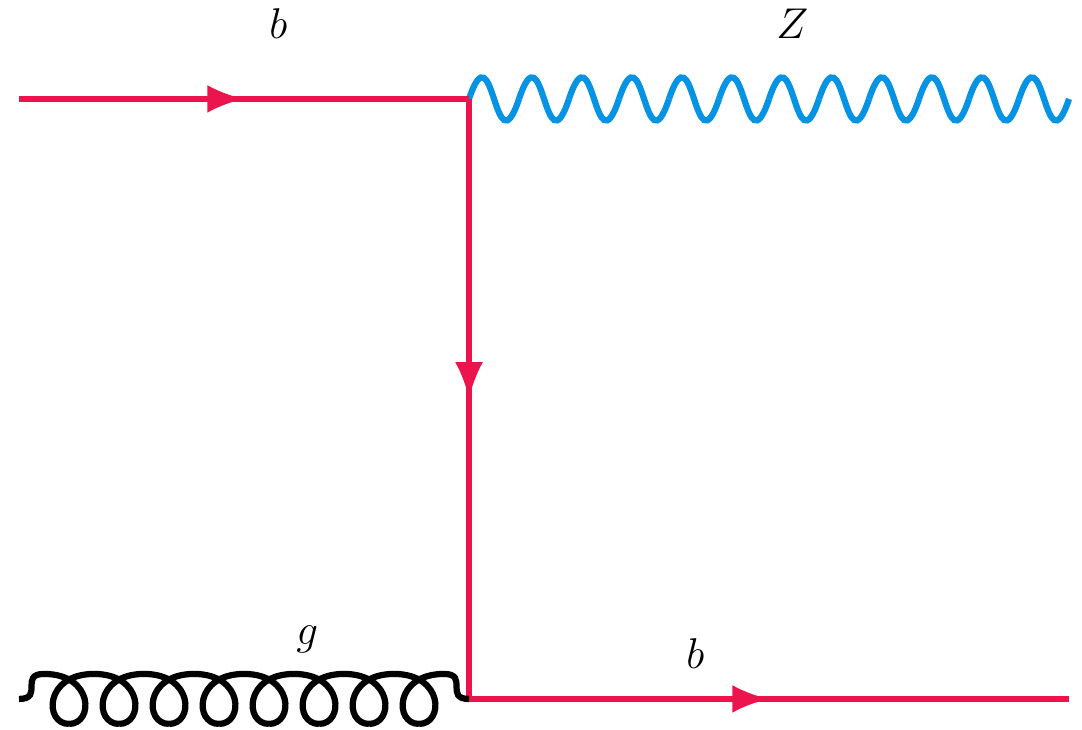, width=0.2\textwidth, clip=}}
  \vspace*{0.2in}
\hspace*{.1in}
  \subfigure[]{\label{fig:process3}
  \epsfig{file=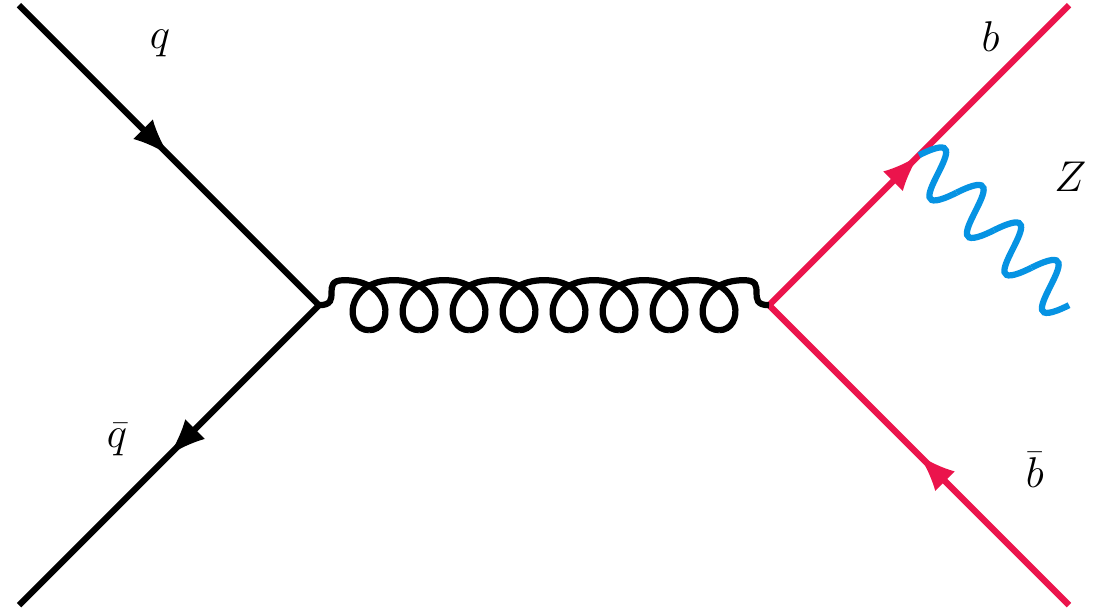, width=0.2\textwidth, clip=}}
  \vspace*{0.2in}
\hspace*{.1in}
 \subfigure[]{\label{fig:process4}
  \epsfig{file=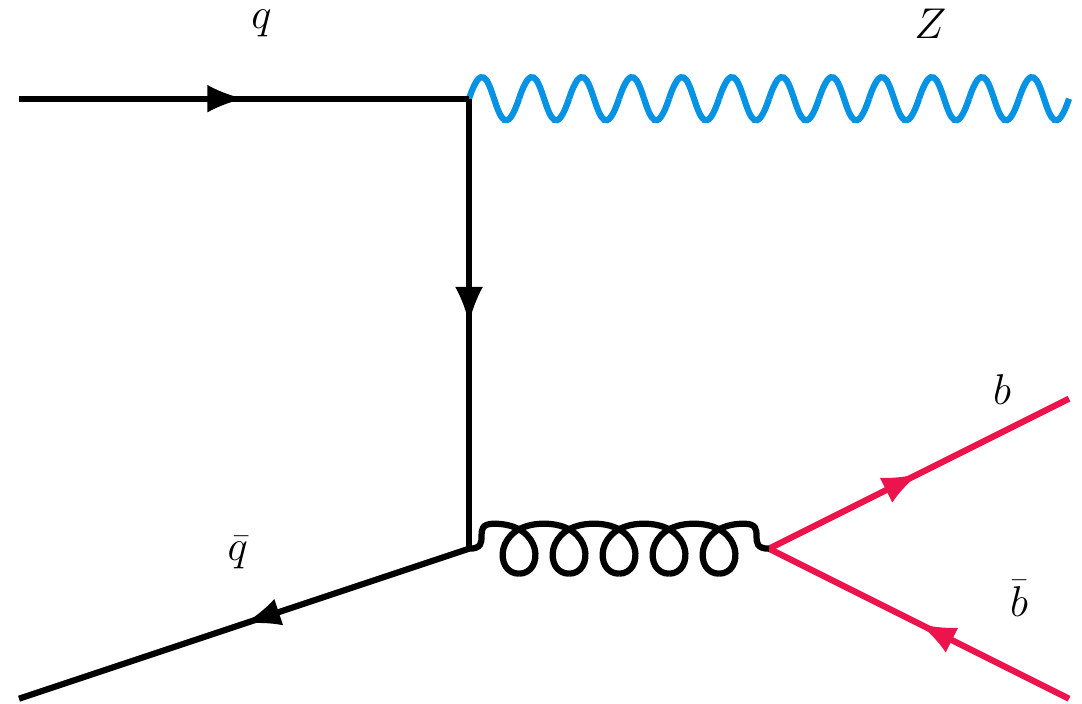, width=0.2\textwidth, clip=}}
  \caption{Feynman diagrams contributing to the associated production of the b-jet with the $Z^0$ boson.}
  \label{fig:process}
\end{center}
\end{figure}

Before we move into the study on $Z^0\,+\,$b-jet production in HIC, we should address its production in p+p collisions. Fig.~\ref{fig:process} shows a few processes~\cite{Aad:2011jn} contributing to the associated production of a $Z^0$ boson and b-jet. In Figs.~\ref{fig:process}(a) and (b), an initial bottom quark from the parton distribution function~(PDF) derived from the gluon distribution of one beam particle suffers the hard scattering and then turns into a b-jet and a $Z^0$ boson in the final state. In the bottom two diagrams, the $b\bar{b}$ pairs originate from the hard scattering and then turn into two b-jets associated with an emitted $Z^0$ boson in the final state.

In this work, we use the MC@NLO event generator SHERPA~\citep{Gleisberg:2008ta} to obtain the initial $Z^0\,+\,$b-jet production in a p+p collision. The tree-level matrix elements are calculated using the internal modules Amegic~\citep{Krauss:2001iv} and Comix~\citep{Gleisberg:2008fv}, and the one-loop virtual correction is calculated using the external program BlackHat~\cite{Berger:2008ag}.  The parton shower based on the Catani-Seymour~\citep{Schumann:2007mg} subtraction method is matched with NLO QCD matrix elements via the MC@NLO method~\citep{Frixione:2002ik}. The NLO PDF from NNPDF3.0~\citep{Ball:2014uwa} with a 5-flavor scheme has been chosen in the calculations. FastJet~\citep{Cacciari:2011ma} with an anti-$k_{\rm T}$ algorithm is used in the final-state jet reconstruction.

To compare our calculation results based on SHERPA with experimental data for p+p collisions, the same configurations implemented by the CMS collaboration~\citep{Khachatryan:2016iob} were used in our simulations. The $Z^0$ boson is reconstructed according to its decay channels $Z^0\rightarrow e^{+}e^{-}$ and $Z^0\rightarrow \mu^{+}\mu^{-}$. The transverse momentum of the electron and muon candidates is required to be larger than 20~GeV. To exclude the barrel-endcap transition region, the electrons are selected within the pseudorapidity region $|\eta|<1.44$ or $1.57<|\eta|<2.4$, while muons are selected within $|\eta|<2.4$. According to the requirement of the experiment, the events are considered only when the invariant mass of the electron or muon pairs lies in the region $70<M_{ll}<111$~GeV. The jets associated with the $Z^0$ boson are reconstructed by FastJet using the anti-$k_T$ algorithm with a cone size of $R=0.5$. To reduce the contribution from the underlying event, the reconstructed jets must be in the pseudorapidity region $|\eta^{\rm jet}|<2.4$ and have $p_{T,\rm jet}>30$~GeV. The contribution from the underlying event is less than $5\%$, as estimated by CMS~\citep{Khachatryan:2016iob} because the production of softer jets is significantly suppressed by this requirement.

\begin{figure}[!t]
\begin{center}
\vspace*{-0.2in}
\hspace*{-.1in}
\includegraphics[width=3in,height=2.8in,angle=0]{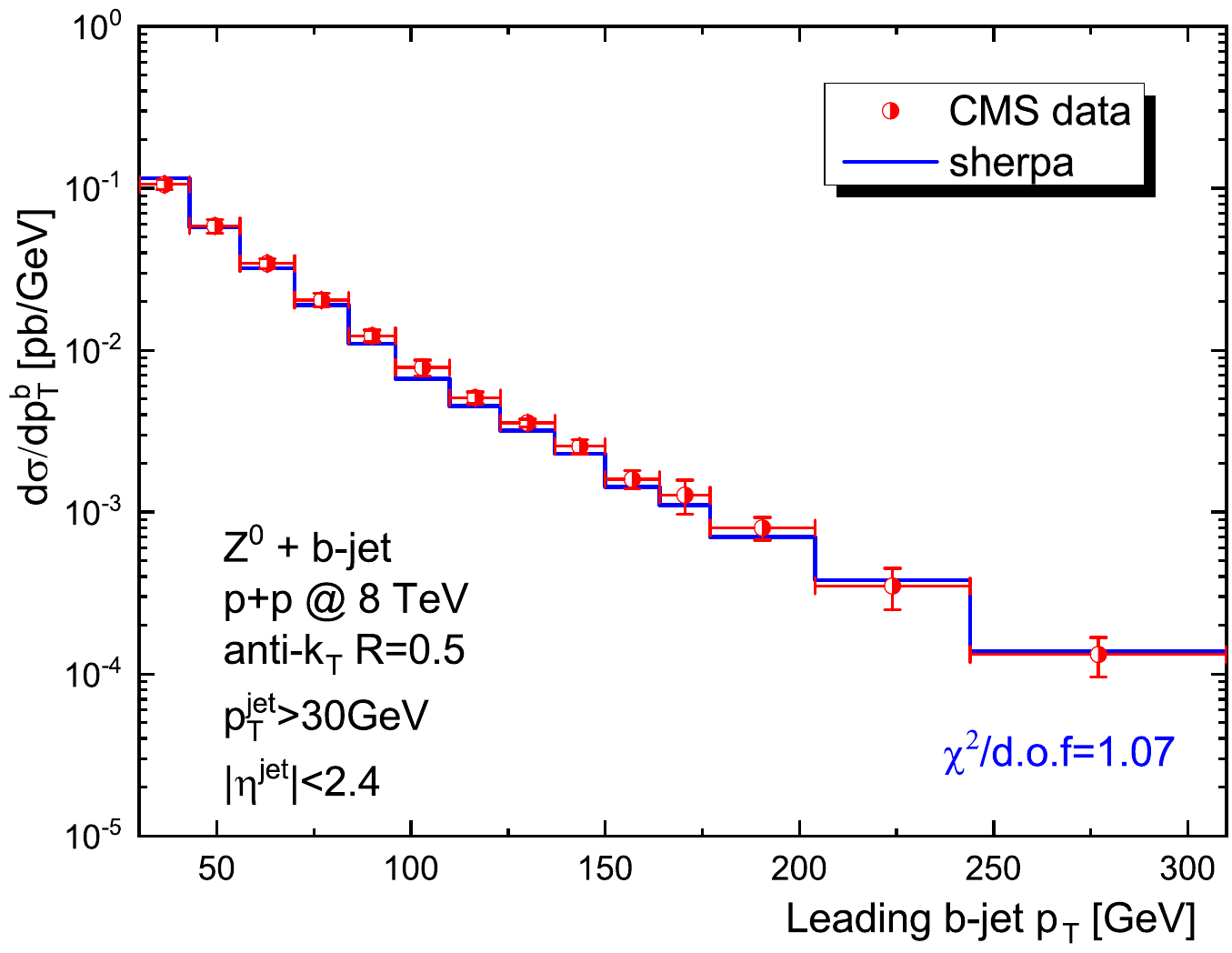}
\includegraphics[width=3in,height=2.8in,angle=0]{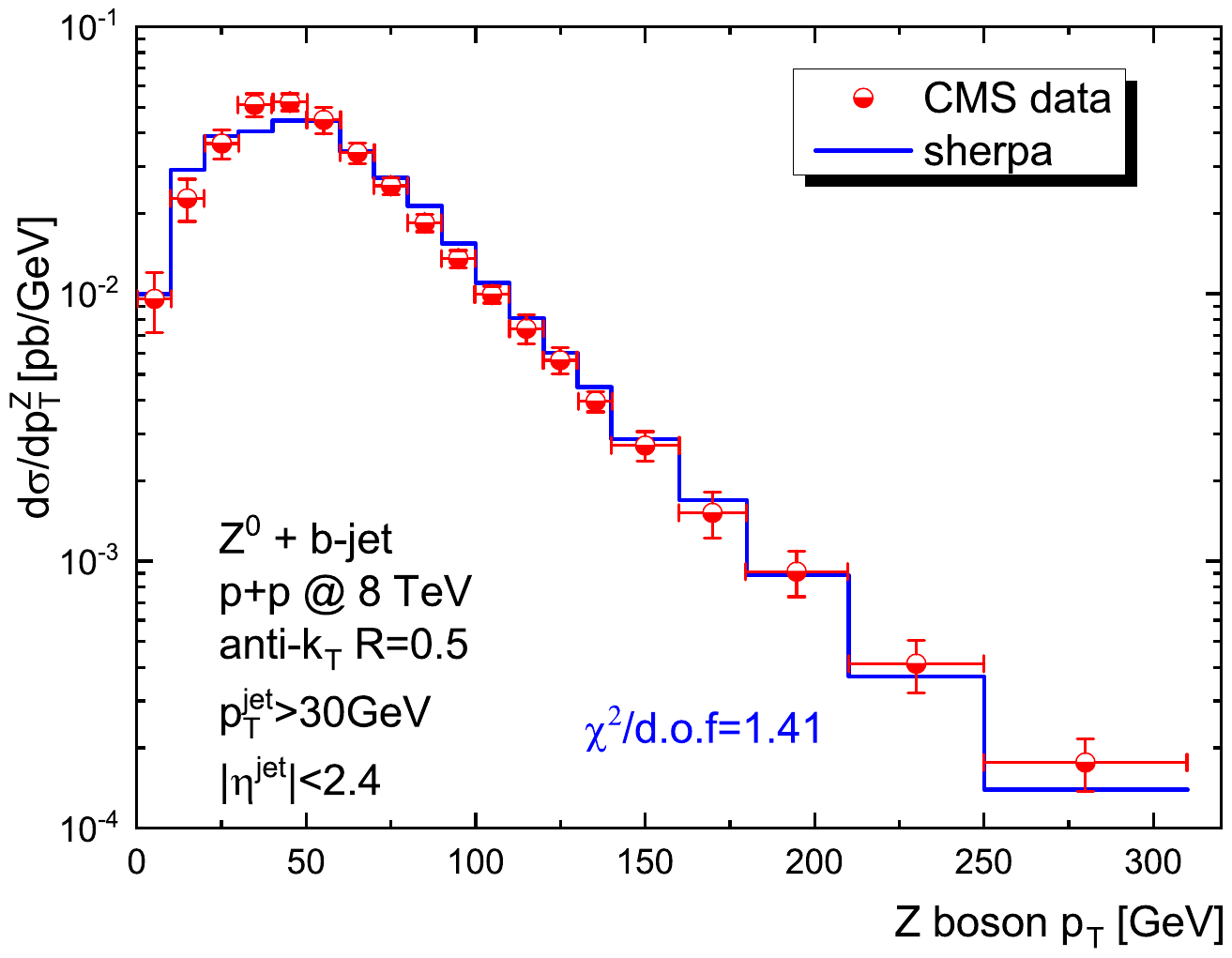}
\vspace*{.1in}
\caption{Differential cross section of $Z^0\,+\,$b-jet simulated by SHERPA~(blue line) in the p+p collision at $\sqrt{s}=8$~TeV  as a function of transverse momentum of the highest-$p_T$ b-jet~(upper panel) and the transverse momentum of the $Z^0$ boson~(bottom panel) compared with the CMS data~\citep{Khachatryan:2016iob}; the $\chi^2/\rm d.o.f$ of our fit to the CMS data is also presented in the plots.}
\label{fig:ptzb}
\end{center}
\end{figure}

\begin{figure}[!t]
\begin{center}
\vspace*{-0.2in}
\hspace*{-.1in}
\includegraphics[width=3in,height=2.8in,angle=0]{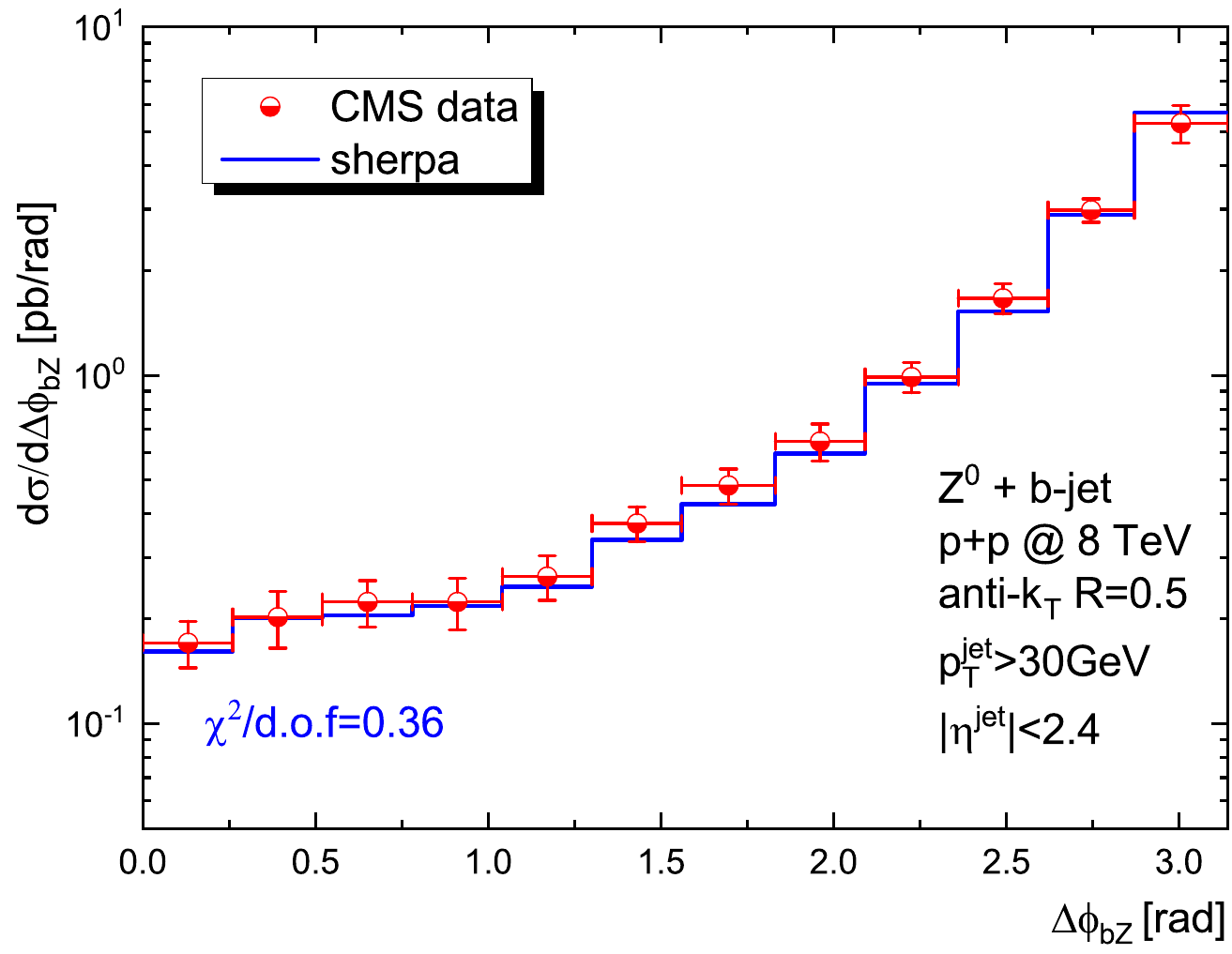}
\includegraphics[width=3in,height=2.8in,angle=0]{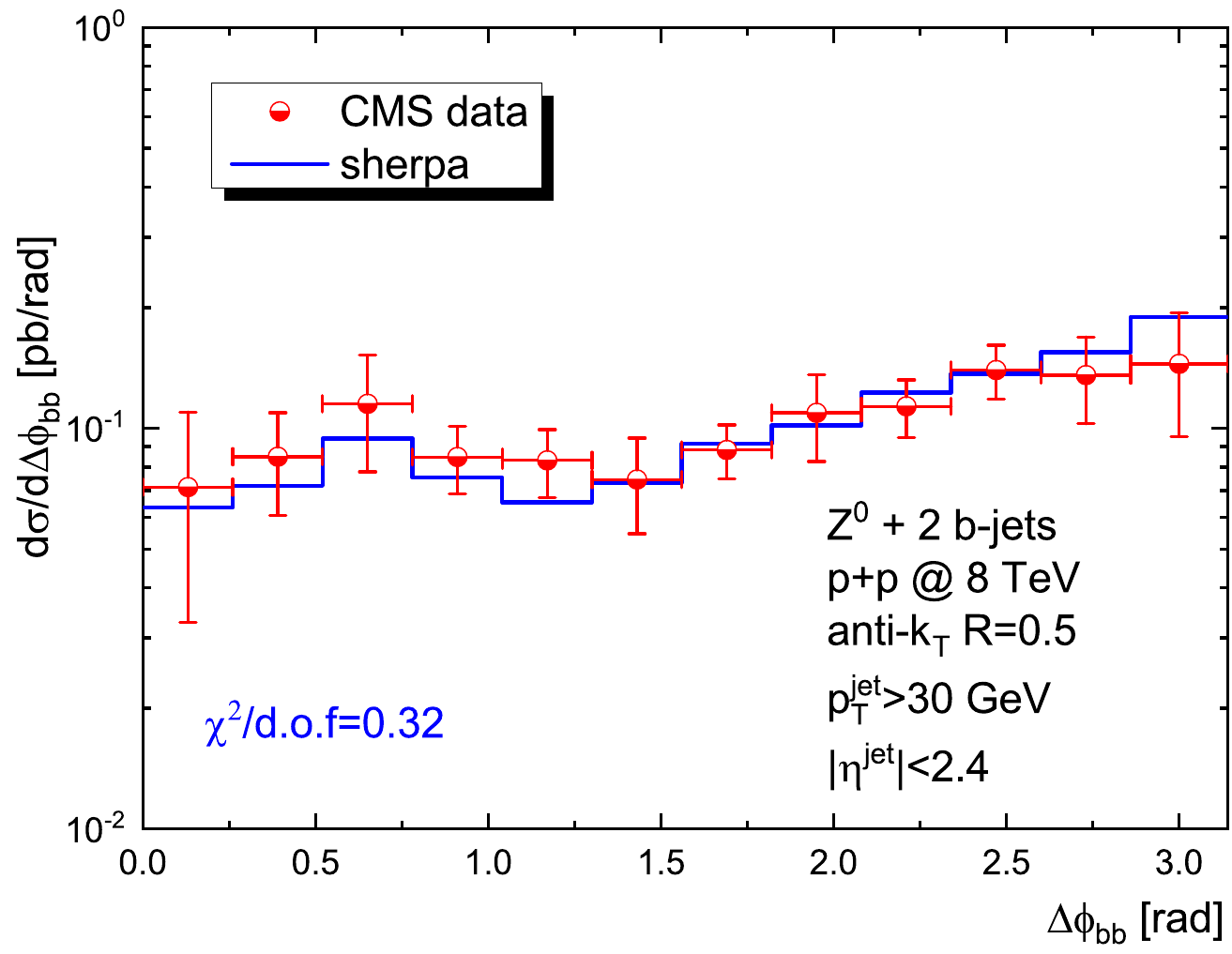}
\vspace*{.1in}
\caption{Differential cross section of $Z^0\,+\,$b-jet simulated by SHERPA~(blue line) in the p+p collision at $\sqrt{s}=8$~TeV as a function of azimuthal angular difference $\Delta\phi_{bZ}=|\phi_{b}-\phi_{Z}|$ of $Z^0$ boson and b-jet~(upper panel) and the azimuthal angular difference $\Delta\phi_{bb}=|\phi_{b1}-\phi_{b2}|$ of the two b-jets~(bottom panel) compared with the CMS data~\cite{Khachatryan:2016iob}; the $\chi^2/\rm d.o.f$ of our fit to the CMS data is also presented in the plots.}
\label{fig:phizb}
\end{center}
\end{figure}

For events with at least one b-jet, we show the differential cross sections calculated via SHERPA as a function of the leading b-jet $p_T$ and $Z^0$ boson $p_T$ in Fig.~\ref{fig:ptzb}. In the upper panel of Fig.~\ref{fig:phizb}, the azimuthal angular correlation between the leading b-jet and $Z^0$ boson~($\Delta\phi_{bZ}=|\phi_{b}-\phi_{Z}|$) is compared between the calculation and the CMS data. In the lower panel of Fig.~\ref{fig:phizb}, for events with two b-jets~(hereinafter $Z^0\,+\,2\,$b-jets), we plot the differential cross sections as a function of the azimuthal angle between the two b-jets~($\Delta\phi_{bb}=|\phi_{b1}-\phi_{b2}|$). To quantify the deviation of our calculations from the experimental data points, we estimate the $\chi^2/\rm d.o.f$ of these observables and present their values in the figures, where $\chi^2=\sum_{i}\frac{[D_i-T_i]^2}{\delta_i^2}$; $D_i$ and $\delta_i$ are the center value and uncertainties, respectively, of the $i-$th experimental data point, $T_i$ represents the theoretical value; and $\rm d.o.f$ represents the number of compared data point. Our calculations based on SHERPA agree well with the experimental measurements.

\section{In-medium jet evolution}
\label{sec:eloss}

In high-energy nuclear collisions, a droplet of an exotic state of nuclear matter, i.e., the QGP, is expected to be formed. The high $p_T$ partons produced in the hard scattering propagating in the QGP suffer both collisional and radiative energy loss as a result of the in-medium interaction. Numerous theoretical approaches ~\cite{vanHees:2007me,CaronHuot:2008uh,vanHees:2005wb,Djordjevic:2015hra,He:2014cla,Chien:2015vja,Kang:2016ofv,Alberico:2013bza} and Monte Claro models~\cite{Cao:2011et,Moore:2004tg,Cao:2015hia,Cao:2017hhk,Xu:2015bbz,Cao:2016gvr,Das:2016cwd,Ke:2018tsh} have been developed in the last two decades to describe the heavy flavor meson production in HIC both at the RHIC and LHC. Among them, the Langevin transport equations have been employed effectively to describe the heavy quark evolution in the expanding QCD medium~\citep{Svetitsky:1987gq,Nahrgang:2013saa,Scardina:2017ipo,Cao:2013ita,He:2011zx,Dai:2018mhw,Wang:2019xey}. Because complete treatment of the heavy quark jets propagating in the QGP usually needs a simultaneous description of space-time evolution for both light and heavy partons~\cite{Rapp:2018qla,Djordjevic:2013xoa,Kang:2016ofv,Sharma:2009hn}, we use the modified Langevin equations~\cite{Cao:2013ita,Dai:2018mhw,Wang:2019xey} to describe the propagation of heavy quarks in QGP and take into account the in-medium energy loss of light partons~\cite{Neufeld:2010xi,Huang:2013vaa,Guo:2000nz,Zhang:2003yn,Zhang:2004qm,Majumder:2009ge}. Additionally, owing the lack of a unified theoretical approach that covers the full phase space of in-medium jet evolution in HIC, there are usually two methods to combine the vacuum parton shower with the medium-induced radiation \cite{Armesto:2011ht,Cao:2017qpx}. The first one is to introduce medium modifications on the vacuum parton splitting at higher virtuality scale $t\ge\hat{q}\tau_f$ (where $\hat{q}$ is the jet transport parameter and $\tau_f$ is the formation length of the radiated gluon), as in the Q-PYTHIA~\cite{Armesto:2009fj}, MATTER~\cite{Majumder:2013re} and JEWEL~\cite{Zapp:2013vla} models. In this work, we employ the alternative treatment implemented in LBT~\cite{He:2015pra} and MARTINI~\cite{Schenke:2009gb} to simulate the jet energy loss at low virtuality $t\le\hat{q}\tau_f$ and high energy $E\gg q_{\perp}$ (where $q_{\perp}$ represents the momentum exchange between the hard parton and the hot medium). According to this strategy, we take the p+p events produced by SHERPA with full vacuum parton shower as the input, sample their initial spatial positions using the MC-Glauber model~\cite{Miller:2007ri}, and then simulate the subsequent in-medium evolution.

\subsection{Collisional energy loss}

The movement of a heavy quark with large mass~($M\gg T$) propagating in the hot/dense nuclear matter and suffering a large number of random kicks from the medium can be modeled as a Brownian motion~\citep{Svetitsky:1987gq}. Hence a discrete Langevin equation can be utilized to describe the propagation of heavy quarks in the QCD medium~\citep{Moore:2004tg, Cao:2013ita,Dai:2018mhw,Wang:2019xey}:

\begin{eqnarray}
&&\vec{x}(t+\Delta t)=\vec{x}(t)+\frac{\vec{p}(t)}{E}\Delta t\\
&&\vec{p}(t+\Delta t)=\vec{p}(t)-\Gamma(p)\vec{p} \Delta t+\vec{\xi}(t)\Delta t-\vec{p}_g \, ,
\label{eq:lang2}
\end{eqnarray}
where $\Delta t$ represents the timestep in the Monte Carlo simulation, and $\Gamma(p) $ is the drag coefficient representing the dissipation effect and controlling the strength of quasi-elastic scattering. $\vec{\xi}(t)$ is the stochastic term that obeys a Gaussian probability distribution,
\begin{eqnarray}
W[\vec{\xi}(t)]=N\exp[\frac{\vec{\xi}(t)^2}{2\kappa/\Delta t}] \, ,
\end{eqnarray}
and leads to

\begin{eqnarray}
&\left\langle \xi_i(t) \right\rangle&=0\\
&\left\langle \xi_i(t)\xi_j(t') \right\rangle&=\kappa\delta_{ij}(t-t')\, ,
\end{eqnarray}

The diffusion coefficient $\kappa$ is related to the drag coefficient $\Gamma$ by the fluctuation-dissipation relation~\cite{Kubo}:
\begin{eqnarray}
\kappa=2\Gamma ET=\frac{2T^2}{D_s} \, ,
\end{eqnarray}
where $D_s$ is the spacial diffusion coefficient. The last term $-\vec{p}_g$ is the recoil momentum due to the medium-induced gluon radiation, which is discussed in the following section. At each timestep, we boost partons to the local rest frame of the expanding medium to update the four-momentum and then boost them back to the laboratory frame to update the spatial position. Note that the procedure is done for $T>T_c$, where $T_c=165$~MeV is the QCD transition temperature~\cite{Aoki:2006br,Cheng:2006qk}. The space-time evolution profile of the bulk medium in Pb+Pb collision is provided by the smoothed VISHNU~\cite{Shen:2014vra} code. Even though the event-by-event fluctuation effects on the jet energy loss are small, the initial geometry fluctuation may be non-negligible for other observables such as particle collective flow ($v_n$)~\cite{Renk:2011qi,Betz:2011tu,Noronha-Hostler:2016eow}.

\par Meanwhile, the calculation for leading logarithmic accuracy at Hard-Thermal-Loop approximation~\cite{Neufeld:2010xi,Huang:2013vaa} is employed in our framework to take into account the collisional energy loss of light quarks and gluons:
\begin{eqnarray}
\frac{dE}{dL}=-\frac{\alpha_{s}C_{s}\mu_{D}^{2}}{2}ln{\frac{\sqrt{ET}}{\mu_D}}
\label{eq:HTL}
\end{eqnarray}
where $L$ represent the transport path of the partons along the propagating direction, $\alpha_s$ is the strong coupling constant, $C_s$ is the quadratic Casimir in color representation, and $\mu_{D}$ represents the Debye screening mass in the QCD medium. Note that Eq.~(\ref{eq:HTL}) is only employed for light partons to consider their collisional energy loss because they cannot be treated as massive particles to evolve in the medium with the Langevin equations. During each timestep, the amount of collisional energy loss of a light parton can be calculated by integrating Eq.~(\ref{eq:HTL}). Because medium-induced gluon radiation is the dominant energy loss mechanism for light partons, this treatment can be regarded as an effective approximation.

\subsection{Medium induced gluon radiation}

\par The inelastic scattering also plays an important role in the in-medium energy loss of energetic partons~\cite{Zakharov:2007pj,Qin:2007rn}. In our work, the Higher-Twist~(HT) radiated gluon spectra~\cite{Guo:2000nz,Zhang:2003yn,Zhang:2004qm,Majumder:2009ge} is implemented to simulate the medium-induced gluon radiation when a parton propagates in the dense and hot QCD matter:

\begin{eqnarray}
\frac{dN}{dxdk^{2}_{\perp}dt}=\frac{2\alpha_{s}C_sP(x)\hat{q}}{\pi k^{4}_{\perp}}\sin^2(\frac{t-t_i}{2\tau_f})(\frac{k^2_{\perp}}{k^2_{\perp}+x^2m^2})^4\nonumber\\
\label{eq:spec}
\end{eqnarray}
where $x$ and $k_\perp$ represent the energy fraction and the transverse momentum of the radiated gluon, respectively. $\alpha_s$ is the strong coupling constant, which is fixed at $\alpha_s=0.3$ in our calculations, $C_s$ is the quadratic Casimir in color representation, and $P(x)$ is splitting function~\cite{Wang:2009qb} for the splitting processes $q\rightarrow q+g$ and $g\rightarrow g+g$ ($g\rightarrow q+\bar{q}$ process is negligible owing to its low probability~\cite{He:2011zx}).

\begin{eqnarray}
P_{q\rightarrow qg}(x)=&\frac{(1-x)(1+(1-x)^2)}{x}\\
P_{g\rightarrow gg}(x)=&\frac{2(1-x+x^2)^3}{x(1-x)}
\end{eqnarray}
$\tau_f$ is the radiated gluon formation time defined as $\tau_f=2Ex(1-x)/(k^2_\perp+x^2m^2)$, and $t-t_i$ is the time interval between two instead of inelastic scattering. In the rigorous Higher-Twist calculations, the splitting function $P(x)$ of heavy quarks should be mass dependent \cite{Zhang:2004qm}. Nevertheless, as a simplified treatment in many transport models, such as LBT \cite{He:2015pra,Cao:2016gvr}, QLBT~\cite{Liu:2021dpm}, and BAMPS \cite{Uphoff:2014hza,Uphoff:2014cba}, it's convenient to uniformly write $P(x)$ as a massless form for heavy and light quarks, while the dominant mass effect of the gluon radiation spectra of heavy quarks can be approximately presented as an overall factor $(k_{\perp}^2/(k_{\perp}^2+x^2m^2))^4$. Additionally, $\hat{q}$ is the jet transport coefficient~\cite{Chen:2010te}:
\begin{eqnarray}
\hat{q}(\tau,\vec{r})=q_0\frac{\rho^{QGP}(\tau,\vec{r})}{\rho^{QGP}(\tau_0,0)}\frac{p^{\mu}u_{\mu}}{p^0}
\end{eqnarray}
where $\hat{q}_{0}$ denotes the value of $\hat{q}$ at the center of the bulk medium at the initial time $\tau_0=0.6~\rm fm/c$, and $\rho^{QGP}(\tau,r)$ is the parton number density where the parton is probed. To take into account the radial flow effect~\cite{Baier:2006pt}, the four-momentum of parton  $p^{\mu}$ and the four flow velocity of the medium in the collision frame $u^{\mu}$ act as a modification for $\hat{q}$ in an expanding nuclear medium. We still hold that $\hat{q}$/$T^3=\rm const.$ in the current framework as an effective approximation. More studies focusing on the puzzle of the temperature dependence of $\hat{q}/T^3$ can be found in the Refs.~\cite{Burke:2013yra,Xu:2015bbz,Andres:2016iys,Kumar:2019uvu,Xie:2022fak}. The last term of Eq.~(\ref{eq:spec}) represents the dead-cone effect~\cite{Armesto:2003jh,Zhang:2004qm} which suppresses the gluon radiation of heavy quarks at a small angle ($\theta\sim M/E$) owing to their large mass. We also noticed the up-to-data development of the HT approach in Ref.~\cite{Zhang:2018kkn} achieved by including both transverse and longitudinal momentum exchanges between hard partons and the QCD medium and will consider their results in our framework in the future.

An imposed cut-off of the radiated gluon energy fraction $x_{\rm min}=\frac{\mu_{D}}{E}$ is taken to avoid the divergence near $x\rightarrow 0$, where $\mu_D^2=4\pi\alpha_s(1+\frac{n_f}{6}) T^2$ is the Debye screening mass induced by the QGP medium.  Therefore, following the method introduced in~\citep{Cao:2016gvr}, one can estimate the mean number of radiated gluons$\left\langle N(t,\Delta t)\right\rangle$ during a timestep $\Delta t$ by integrating the phase space of $x$, $k_{\perp}$, and $t$ in Eq.~(\ref{eq:spec}):

\begin{eqnarray}
\left\langle N(t,\Delta t)\right\rangle =\int_{t}^{t+\Delta t} dt\int_{x_{\rm min}}^{1} dx\int_{0}^{(xE)^2} dk^2_{\perp}\frac{dN}{dxdk^2_{\perp}dt}\nonumber\\
\end{eqnarray}

By assuming that the multiple gluon radiation is a Poisson process, we obtain the probability distribution of the radiation number $P(n,t,\Delta t)$ during a time step, as well as the total inelastic scattering probability $P_{rad}(t,\Delta t)$:

\begin{eqnarray}
&P(n,t,\Delta t)=\frac{{\left\langle N(t,\Delta t)\right\rangle}^n}{n!}e^{-\left\langle N(t,\Delta t)\right\rangle}\\
&P_{rad}(t,\Delta t)=1-e^{-\left\langle N(t,\Delta t)\right\rangle}
\end{eqnarray}

\begin{figure}[!t]
\begin{center}
\vspace*{-0.2in}
\hspace*{-.1in}
\subfigure[]{\label{fig:xji}
  \epsfig{file=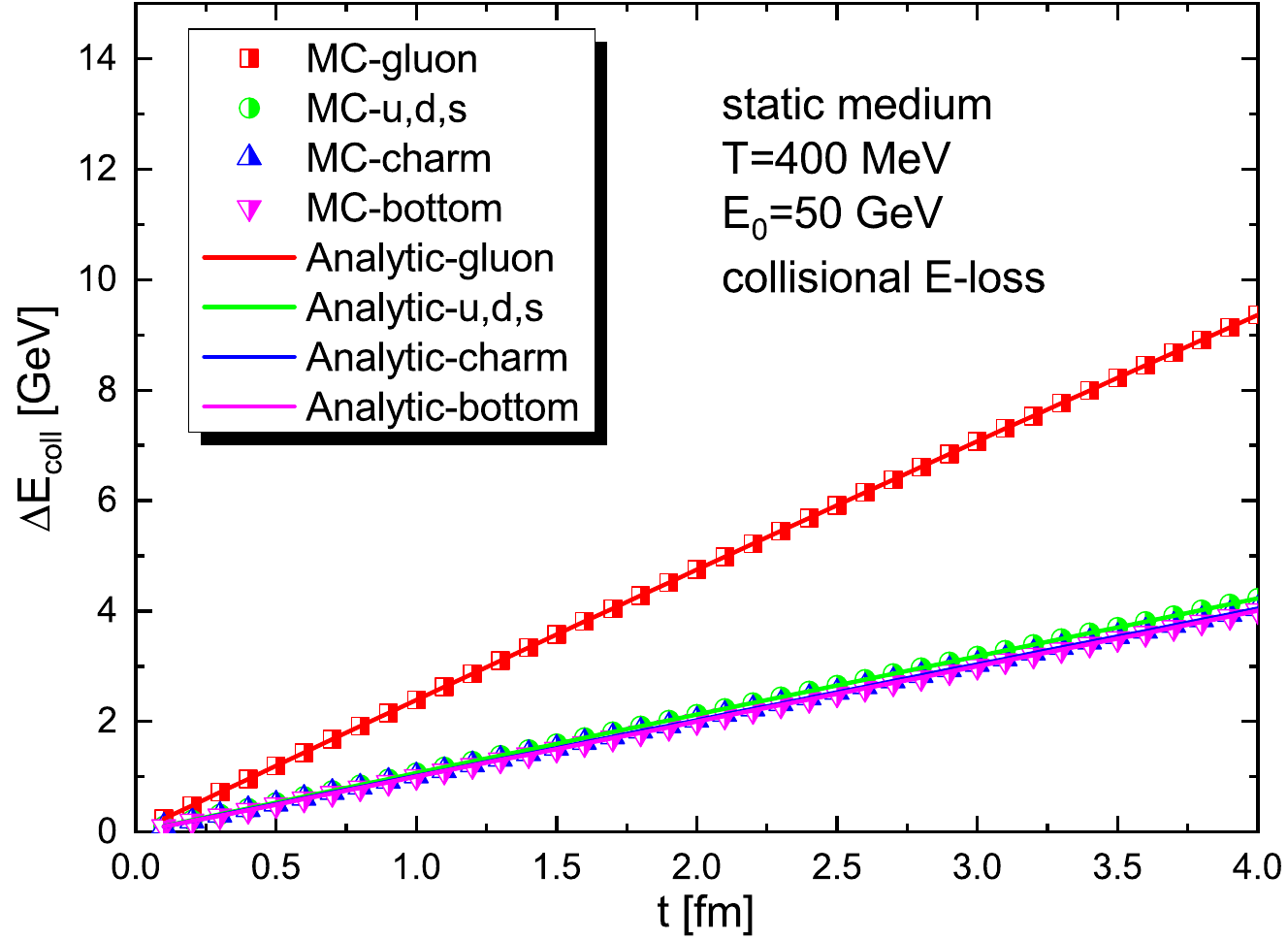, width=0.45\textwidth, clip=}}
  \subfigure[]{\label{fig:xjb}
  \epsfig{file=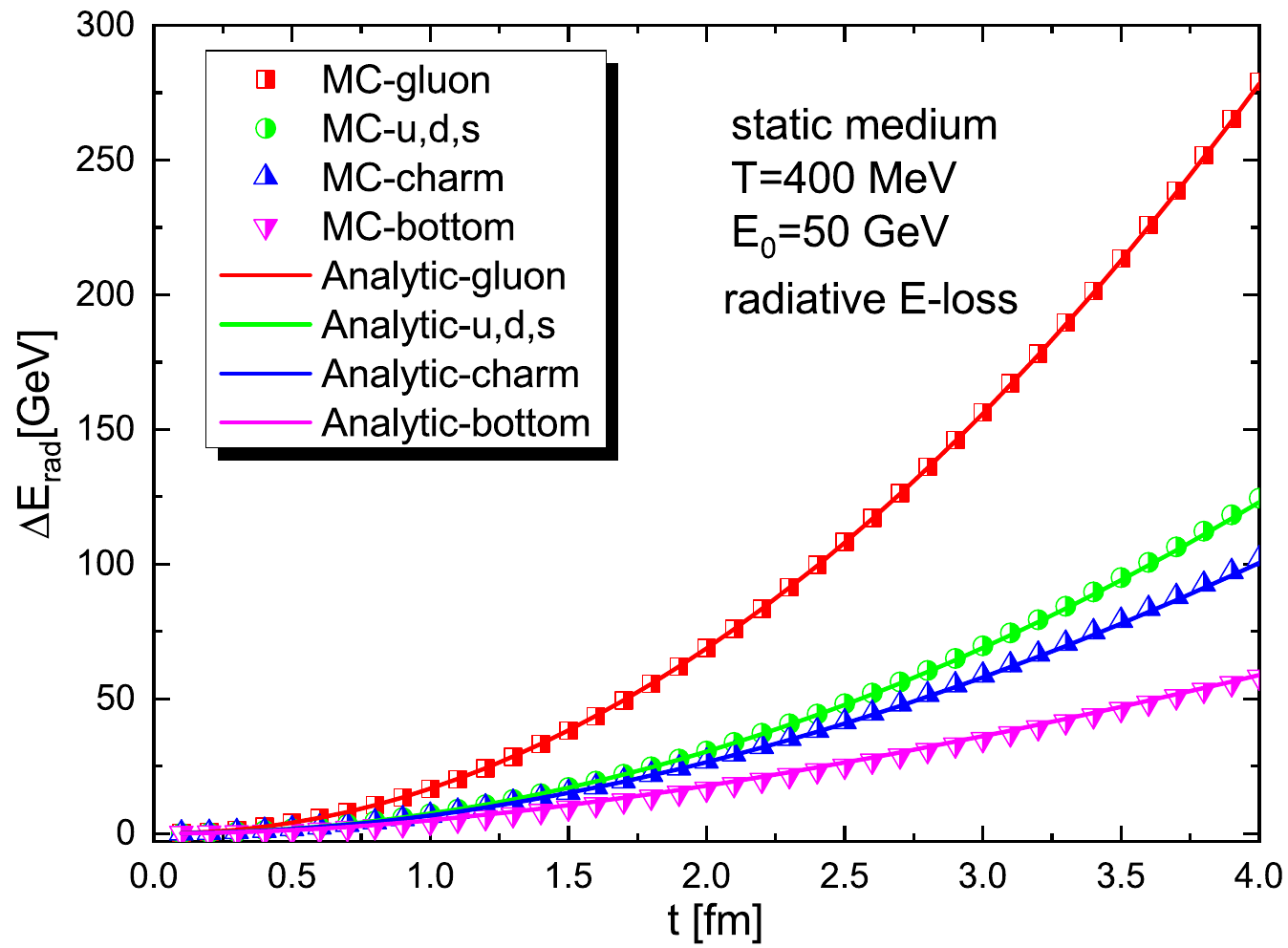, width=0.45\textwidth, clip=}}
\vspace*{0.1in}
\caption{Collisional (a) and radiative (b) energy losses of gluons, light quarks, charm, and bottom with initial energy $E_0=50~$GeV in a static medium with temperature $T=400~$MeV. The Monte Carlo simulations are compared with the semi-analytical calculations.}
\label{fig:rad-mc-anl}
\end{center}
\end{figure}

In our Monte Carlo simulation, during every time step, the $P_{rad}(t,\Delta t)$ is first evaluated to determine whether the radiation occurs. If accepted, the Possion distribution function $P(n,t,\Delta t)$ is used for the sampling of the radiated gluon number. Finally, the four-momentum of the radiated gluon can be sampled according to the spectrum $dN/dxdk_{\perp}^{2}$ expressed in Eq.~(\ref{eq:spec}). In Fig.~\ref{fig:rad-mc-anl}, for a consistent comparison between our Monte Carlo simulation and the analytical calculation, we estimate the collisional and radiative energy losses of the gluons, light quarks, charm, and bottom in a static medium ~($T=400$~MeV). Here we fix the parton energy~(50 GeV) at each evolution time step and also restore the initial time $t_i$ in Eq.~(\ref{eq:spec}) to be 0 (same as the treatment in Ref.~\cite{Cao:2017hhk}) because their variations during the Monte Carlo simulation are not automatically included in the analytical calculation. We find that the MC results agree well with the analytical calculations. For the collisional energy loss, the $\Delta E_{coll}$ of the gluon is $9/4$ times that of the light quark owing to the large color factor, and in our framework, the $\Delta E_{coll}$ of the heavy quark is comparable to that of the light quark. For the radiative energy loss, a clear mass hierarchy for different parton species can be found: $\Delta E^g_{rad}>\Delta E^q_{rad}>\Delta E^c_{rad}>\Delta E^b_{rad}$. For a long propagation time $t=4$~fm, we find that the radiative energy loss dominates the total parton energy loss because of its quadratic dependence on the path length.

In general, there are two parameters in our framework that need to be determined; the jet transport coefficient $\hat{q}$ and the diffusion coefficient $D_s$. We treat $\hat{q}$ and $\kappa$ as two independent parameters to be constrained by experimental data. First, the value of $\hat{q}$ is determined via a global extraction of the single hadron production in Pb+Pb collisions~\cite{Ma:2018swx}, in which $q_0=1.2$~GeV$^2$/fm is obtained at the LHC energy. After $\hat{q}$ is fixed, we extract the best value $D_s(2\pi T)\sim4$  via a $\chi^2$ fitting to the D meson $R_{AA}$ data~\cite{ALICE:2018lyv,Sirunyan:2017xss}, which is consistent with the results of $D_s(2\pi T)=3.7\sim7$ reported by the Lattice QCD~\cite{Banerjee:2011ra}.

\section{Numerical results and discussions}
\label{sec:results}

In this section, to estimate the medium modification of jet observables in nucleus-nucleus collisions, we use the p+p events provided by SHERPA as the input of our simulation within the hydrodynamic background to study the in-medium jet evolution. Before proceeding to the $Z^0$ tagged b-jet, we calculate the azimuthal angular correlation ~($\Delta\phi_{jZ}=|\phi_{\rm jet}-\phi_Z|$) and transverse momentum imbalance~($x_{jZ}=p_T^{\rm jet}/p_T^Z$) of $Z^0\,+\,$jet, as well as the nuclear modification factor $R_{AA}$ of the inclusive b-jet, and compare our theoretical results with the available experimental data.  Then, we calculate the $Z^0\,+\,$b-jet observables, including the azimuthal angular correlation between the $Z^0$ boson and b-jet~($\Delta\phi_{bZ}=|\phi_{\textit{\rm b-jet}}-\phi_Z|$), angle separation between the two Z-tagged b-jets~($\Delta\phi_{bb}=|\phi_{b1}-\phi_{b2}|$), transverse momentum~($x_{bZ}=p_T^{\textit{\rm b-jet}}/p_T^Z$), and nuclear modification factor $I_{AA}$~\cite{Neufeld:2010fj,Kang:2017xnc} defined as

\begin{eqnarray}
I_{AA}=\frac{1}{\left\langle N_{bin} \right\rangle}\frac{\frac{dN^{AA}}{dp^{\rm jet}_{T}}|_{p_T^{\rm min}<p_T^Z<p_T^{\rm max}}}{\frac{dN^{pp}}{dp^{\rm jet}_{T}}|_{p_T^{\rm min}<p_T^Z<p_T^{\rm max}}}
\label{eq:iaa}
\end{eqnarray}
Here, $\left\langle N_{\rm bin} \right\rangle$ denotes the average number of binary nucleon-nucleon collisions among A+A collisions calculated via the Glauber model~\cite{Miller:2007ri}.

\begin{figure}[!t]
\begin{center}
\vspace*{-0.2in}
\hspace*{-.1in}
\includegraphics[width=3in,height=3.8in,angle=0]{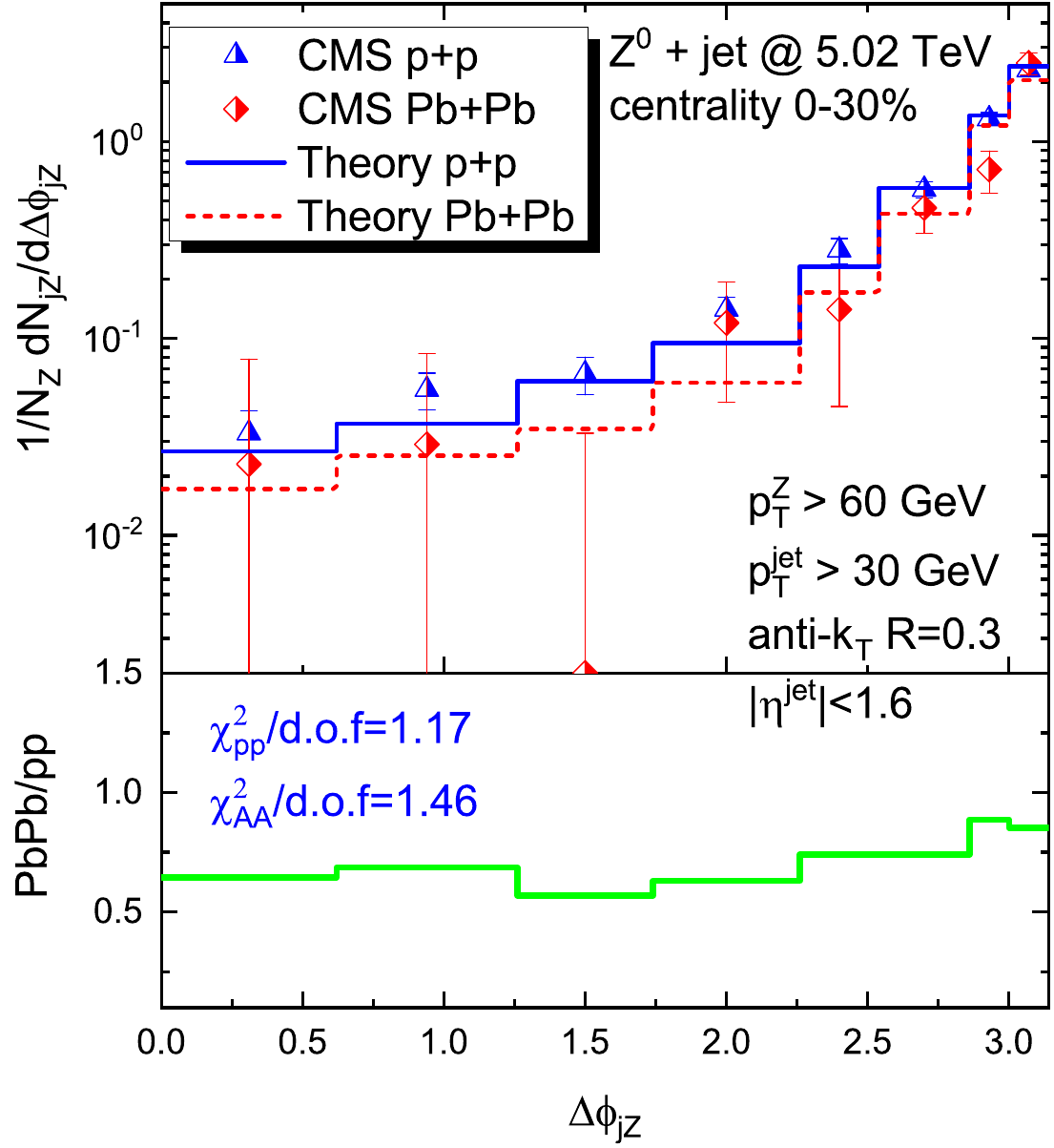}
\vspace*{0.1in}
\caption{Distributions of the azimuthal angle difference $\Delta\phi_{jZ}$ between the $Z^0$ boson and the jet both in p+p and $0-30\%$ Pb+Pb collisions at 5.02~TeV compared with CMS data~\cite{Sirunyan:2017jic}. The distributions are scaled by the number of $Z^0$ events $N$ in p+p collisions. The values of $\chi^2/\rm d.o.f$ of our fit to the CMS data in both p+p and Pb+Pb are also presented in the plots.}
\label{fig:CMSphi}
\end{center}
\end{figure}

In Fig.~\ref{fig:CMSphi}, we show our calculated $\Delta\phi_{jZ}$ distributions both in p+p and $0-30\%$ Pb+Pb collisions at $\sqrt{s_{NN}}=$5.02~TeV compared with the CMS experimental data~\cite{Sirunyan:2017jic}. Additionally, $\chi^2/\rm d.o.f$ of our model fitting to the CMS data both in p+p and Pb+Pb are presented. The same configurations in the jet reconstruction used by the CMS are employed. All final-state jets are reconstructed by FastJet using the anti-$k_T$ algorithm with $R=0.3$ and require $p_T^{\rm jet}>30$~GeV. The selected $Z^0$ boson are reconstructed by the electron or muon pairs based on their decay channels~($Z^0\rightarrow e^{+}e^{-}$ and $Z^0\rightarrow \mu^{+}\mu^{-}$) and require $p_T^{Z}>60$~GeV. Note that these distributions are normalized by the number of $Z^0$ events, and the transverse momentum imbalances are subjected to Gaussian smearing~\cite{Sirunyan:2017jic} to take into account the detector resolution effects.
 This reveals that the distribution of azimuthal angular correlation in Pb+Pb collisions suffers a suppression in a small $\Delta\phi_{jZ}$ region relative to the p+p baseline, which is consistent with the CMS measurement. However, in a large angle region~($\Delta\phi_{jZ}\sim \pi$, where the $Z^0$ boson and jet are almost back-to-back), this suppression is not very apparent. The reason for this behavior has been discussed in detail~\cite{Chen:2018fqu,Luo:2018pto}; i.e., the small $\Delta\phi_{jZ}$ region is dominated by the multiple jets processes and the large $\Delta\phi_{jZ}$ region is dominated by soft/collinear radiation. Usually, the jet energy of multiple-jet processes is relatively low and easier to be shifted below the jet selection threshold~(30~GeV) because of parton energy loss~\cite{Zhang:2018urd}. We also notice two significant differences between the p+p and Pb+Pb CMS data near $\Delta\phi_{jZ}=\pi/2$ and $\Delta\phi_{jZ}=7\pi/8$, and our results cannot fit these points well. However, similar abnormal behaviors of the CMS data are not found in the measurements of the $\gamma-$jet~\cite{Sirunyan:2017qhf}; hence, we guess that they may be caused by the statistical fluctuations in the experiment.

\begin{figure}[!t]
\begin{center}
\vspace*{-0.2in}
\hspace*{-.1in}
\includegraphics[width=3in,height=2.8in,angle=0]{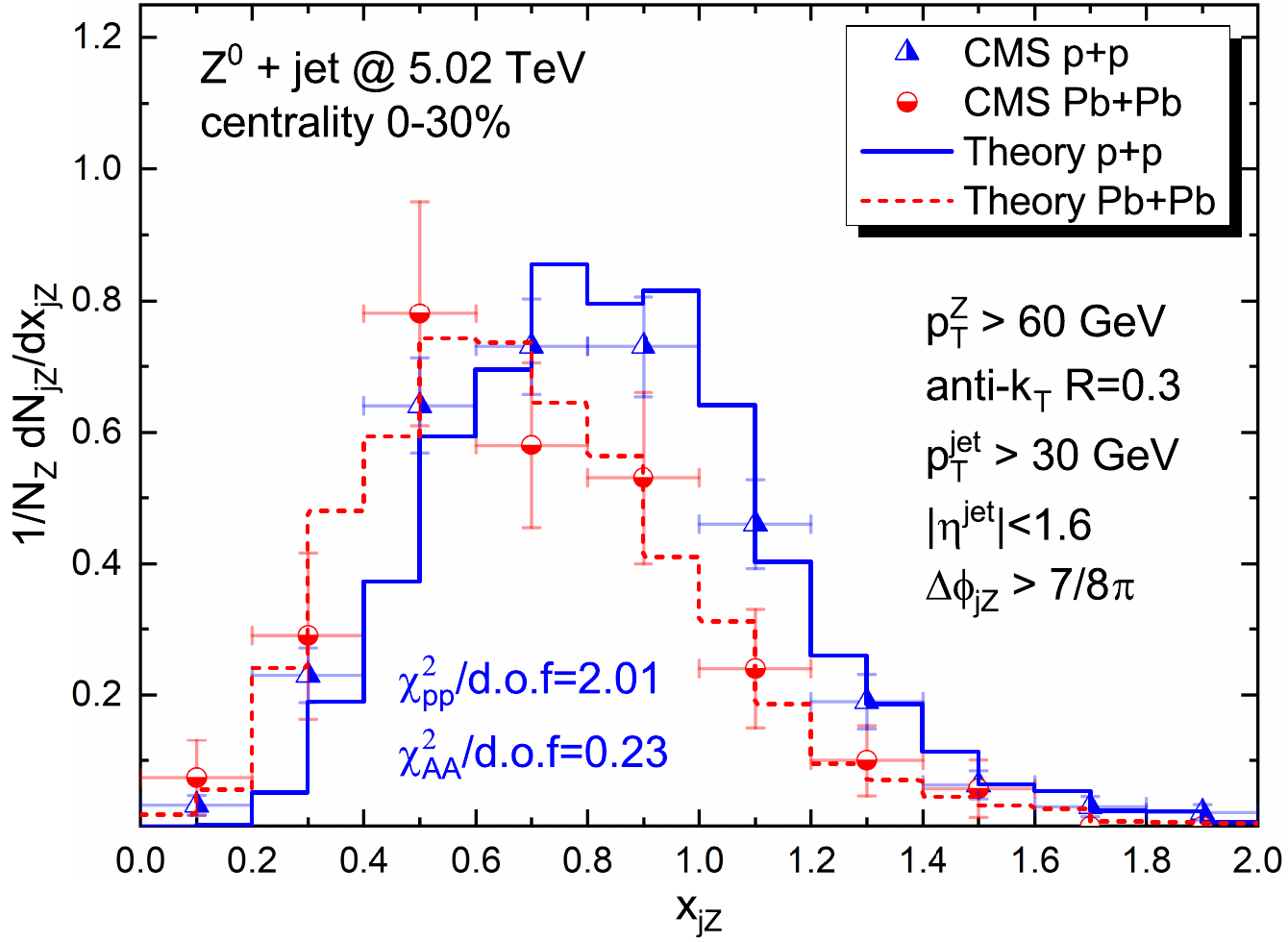}
\vspace*{0.1in}
\caption{Distributions of the transverse momentum balance $x_{jZ}$ of $Z^0\,+\,$jet in both p+p and $0-30\%$ Pb+Pb collisions at 5.02~TeV compared with CMS data~\cite{Sirunyan:2017jic}. The distribution is normalized by the number of $Z^0$ events and $Z^0\,+\,$jet pairs are required with $\Delta\phi_{jZ}>7\pi/8$. The values of $\chi^2/\rm d.o.f$ of our fit to the CMS data for both p+p and Pb+Pb are also presented in the plots.}
\label{fig:CMSxj}
\end{center}
\end{figure}

In Fig.~\ref{fig:CMSxj}, we compute $x_{jZ}$ distribution for $Z^0\,+\,$jet in both p+p and $0-30\%$ Pb+Pb collisions at $\sqrt{s_{NN}}=$5.02 TeV compared with the CMS data. The $\chi^2/\rm d.o.f$ are also presented in the figures indicating that our calculations are consistent with the experimental data, but the p+p baseline needs to be improved. Note here that selected $Z^0\,+\,$jet pairs are required to be almost back-to-back~($\Delta\phi_{jZ}>7\pi/8$). Relative to the p+p baseline, in Pb+Pb collisions, we find that the $x_{jZ}$ distribution is shifted toward smaller values, exhibiting an enhancement at $0<x_{jZ}<0.7$ and suppression at $0.7<x_{jZ}<2$. As $x_{jZ}$ represents the transverse momentum imbalance of $Z^0$ and jet, that for each $Z^0\,+\,$jet pair, the values of $x_{jZ}$ decrease owing to the jet energy loss and thus are shifted to a smaller $x_{jZ}$ observed in the final state.

\begin{figure}[!t]
\begin{center}
\vspace*{-0.2in}
\hspace*{-.1in}
\includegraphics[width=2.8in,height=4in,angle=0]{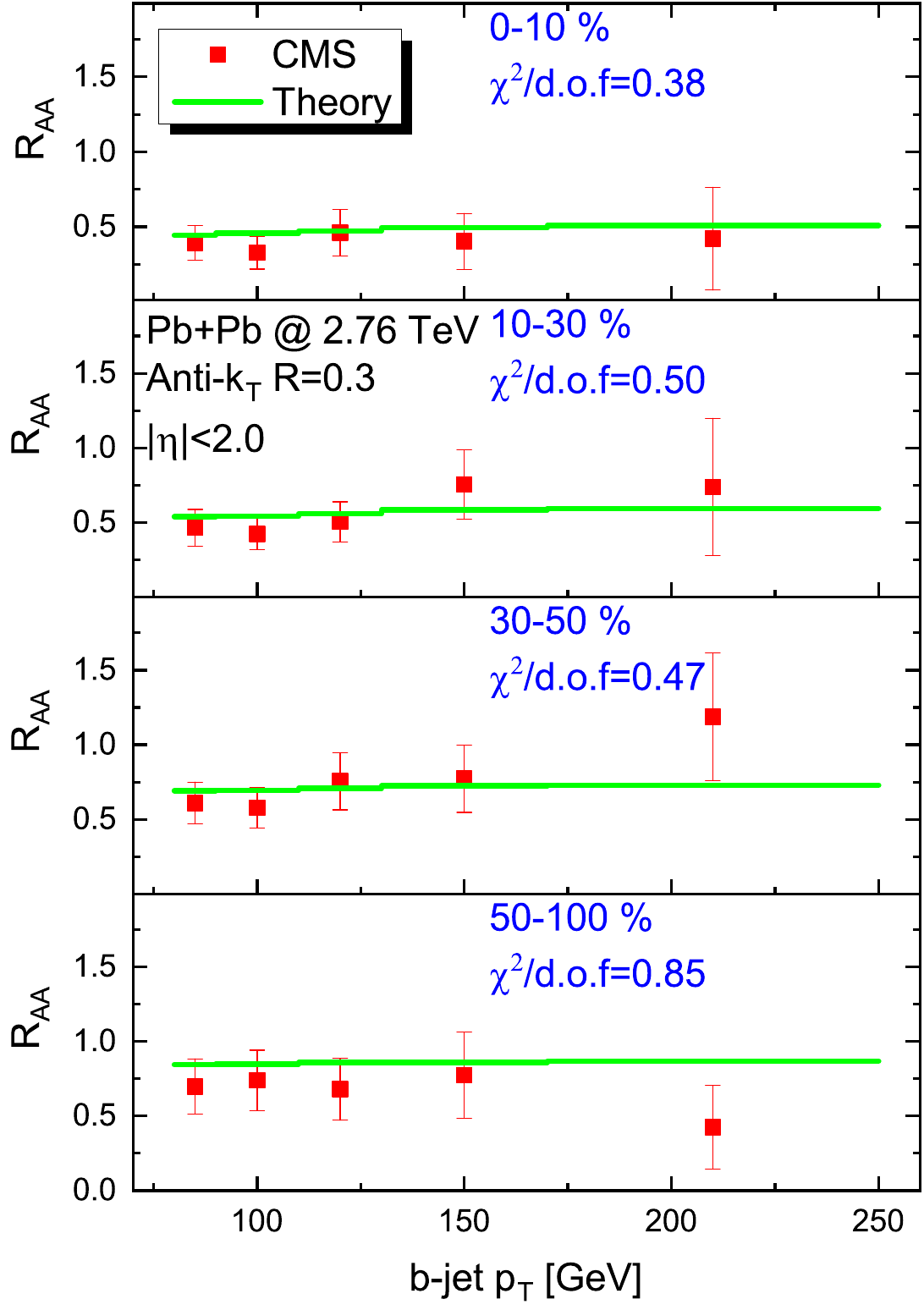}
\vspace*{0.2in}
\caption{Nuclear modification factor $R_{AA}$ of the b-jet. The p+p baseline is provided by SHERPA, and the theoretical calculations are compared with CMS data~\cite{Chatrchyan:2013exa} at centralities of $0-10\%$, $10-30\%$, $30-50\%$, $50-100\%$. The values of $\chi^2/\rm d.o.f$ of our fit to the CMS data are presented in the plots.}
  \label{fig:bjetraa}
\end{center}
\end{figure}

In Fig.~\ref{fig:bjetraa}, we investigate the nuclear modification factor $R_{AA}$ of the inclusive b-jet in Pb+Pb collisions at $\sqrt{s_{NN}}=2.76$~TeV in comparison with the CMS measurements~\cite{Chatrchyan:2013exa} to test our model calculations. The values of $\chi^2/\rm d.o.f$ are shown in each panel of Fig.~\ref{fig:bjetraa} to quantify the deviation of our calculations from the CMS data for different centrality bins. We find that our calculations are essentially consistent with the experimental data, but we may pay attention to some discrepancies. First, our theoretical results show weak $p_T$ dependence but the experimental $R_{AA}$ appears to increase visibly with the jet $p_T$ in the $10-30\%$ and $30-50\%$ centrality bins. Second, some data points ($p_T=100$~GeV at $0-10\%$ and $10-30\%$, $p_T=210$~GeV at $30-50\%$ and $50-100\%$) cannot be well described by our calculations.

\begin{figure}[!t]
\begin{center}
\vspace*{-0.2in}
\hspace*{-.1in}
\includegraphics[width=3in,height=3.5in,angle=0]{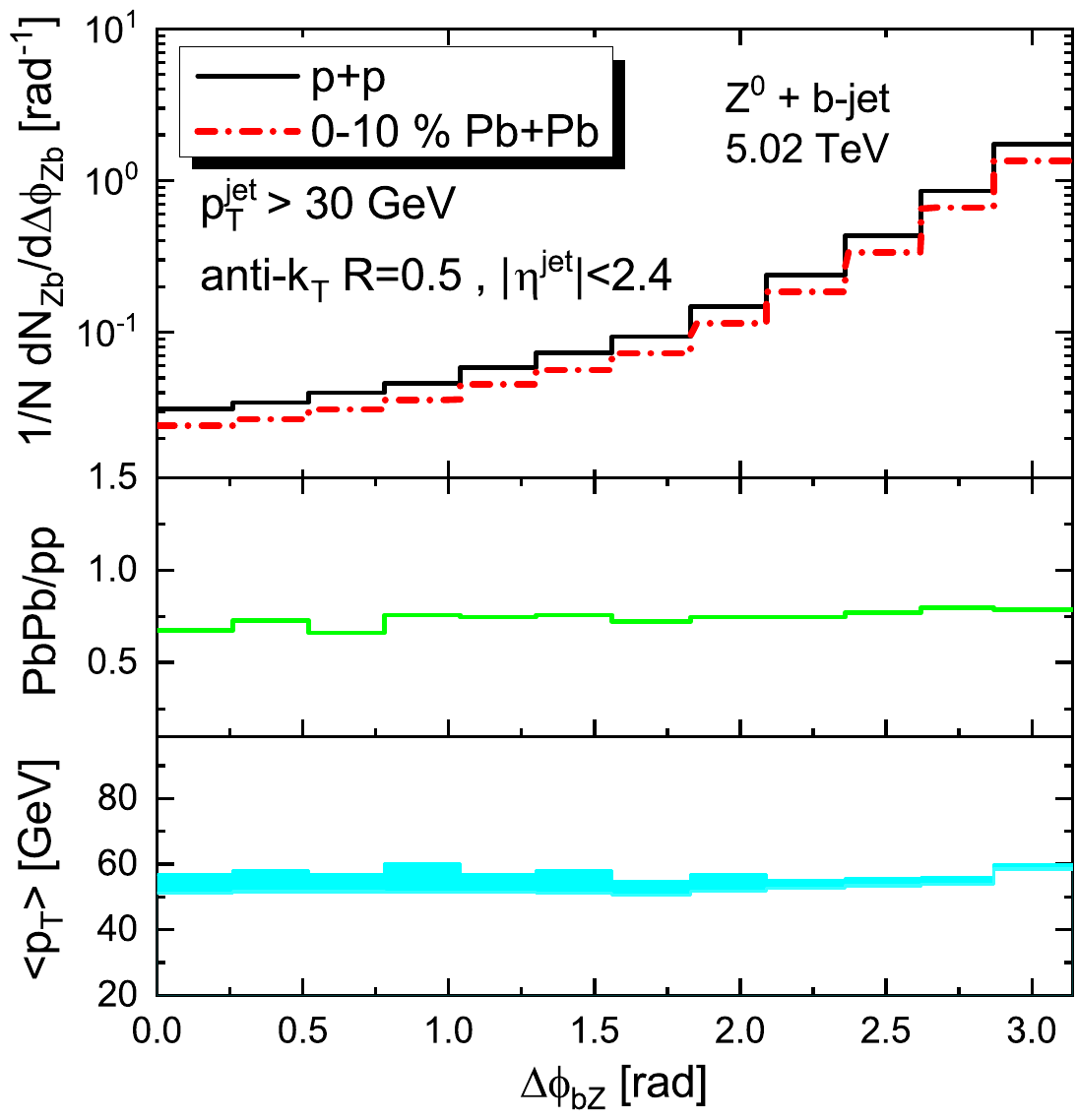}
\vspace*{0.2in}
\caption{Upper panel: distributions of the azimuthal angular correlation of the $Z^0$ boson and b-jet in both p+p and $0-10\%$ Pb+Pb collisions at 5.02~TeV; the distributions are scaled by the number of $Z^0\,+\,$b-jet event $N$ in p+p collisions. Middle panel: ratio of the azimuthal angle correlations in Pb+Pb to p+p. Bottom panel: averaged b-jet transverse momentum $\left\langle p_{T} \right\rangle$ as a function of $\Delta\phi_{bZ}$.
}
  \label{fig:phizbaa}
\end{center}
\end{figure}

The good agreement between our model calculations and the data of the $Z^0\,+\,$jet and inclusive b-jet makes it possible to study the medium modification of $Z^0\,+\,$b-jet in nuclear-nuclear collisions. In Fig.~\ref{fig:phizbaa}, we calculate the azimuthal angular correlation of the $Z^0$ boson and the b-jet in p+p and $0-10\%$ Pb+Pb collisions at $\sqrt{s_{NN}}=$5.02~TeV. The b-jets associated with the $Z^0$ boson are reconstructed by FastJet using the anti-$k_T$ algorithm with a cone size of $\Delta R=0.5$, $|\eta^{\rm jet}|<2.4$ and $p_T^{\rm jet}>30$~GeV for both p+p and Pb+Pb collisions. Note that these distributions are normalized by the initial $Z^0\,+\,$b-jet event number~(in p+p collision) to address the medium modification.
We observe an overall suppression in Pb+Pb collisions relative to the p+p baseline. We show their ratio PbPb/pp in the middle panel of Fig.~\ref{fig:phizbaa}, and find that the suppression for the $Z^0\,+\,$b-jet has a far weaker dependence on $\Delta\phi_{bZ}$ than that for the $Z^0\,+\,$jet, which exhibits stronger suppression in a small $\Delta\phi_{jZ}$ region where multiple jets dominate. All the selected jets must first be b quark tagged, and this requirement significantly reduces the contribution from multiple-jet processes when we consider the azimuthal angular~($\Delta\phi_{bZ}$) distribution.

To address the key factor that leads to the flat suppression on $\Delta\phi_{bZ}$ distribution, we estimate the averaged b-jet transverse momentum $\left\langle p_{T} \right\rangle$ as a function of $\Delta\phi_{bZ}$, which can be calculated by:

\begin{eqnarray}
\left\langle p_{T} \right\rangle (\Delta \phi) = \frac{\int \frac{d\sigma}{dp_Td\Delta\phi}p_Tdp_T}{\int \frac{d\sigma}{dp_Td\Delta\phi}dp_T} \, .
\end{eqnarray}

The decrease of the selected event number in A+A collisions results from the in-medium energy loss, which shifts lower $p_T$ jet below the kinematic selection cut. The initial $\left\langle p_{T} \right\rangle$ distribution actually reflects the $\Delta\phi_{bZ}$ dependence of this shift. It turns out that the distribution of $\left\langle p_{T} \right\rangle$ versus $\Delta\phi_{bZ}$ is nearly a constant value at 55~GeV as shown in the bottom panel of Fig.~\ref{fig:phizbaa}, which leads to the rather flat suppression on $\Delta\phi_{bZ}$ distribution. Here the band of $\left\langle p_{T} \right\rangle$ distribution represents the statistical standard errors in the simulations.

\begin{figure}[!t]
\begin{center}
\vspace*{-0.2in}
\hspace*{-.1in}
\includegraphics[width=3in,height=3.5in,angle=0]{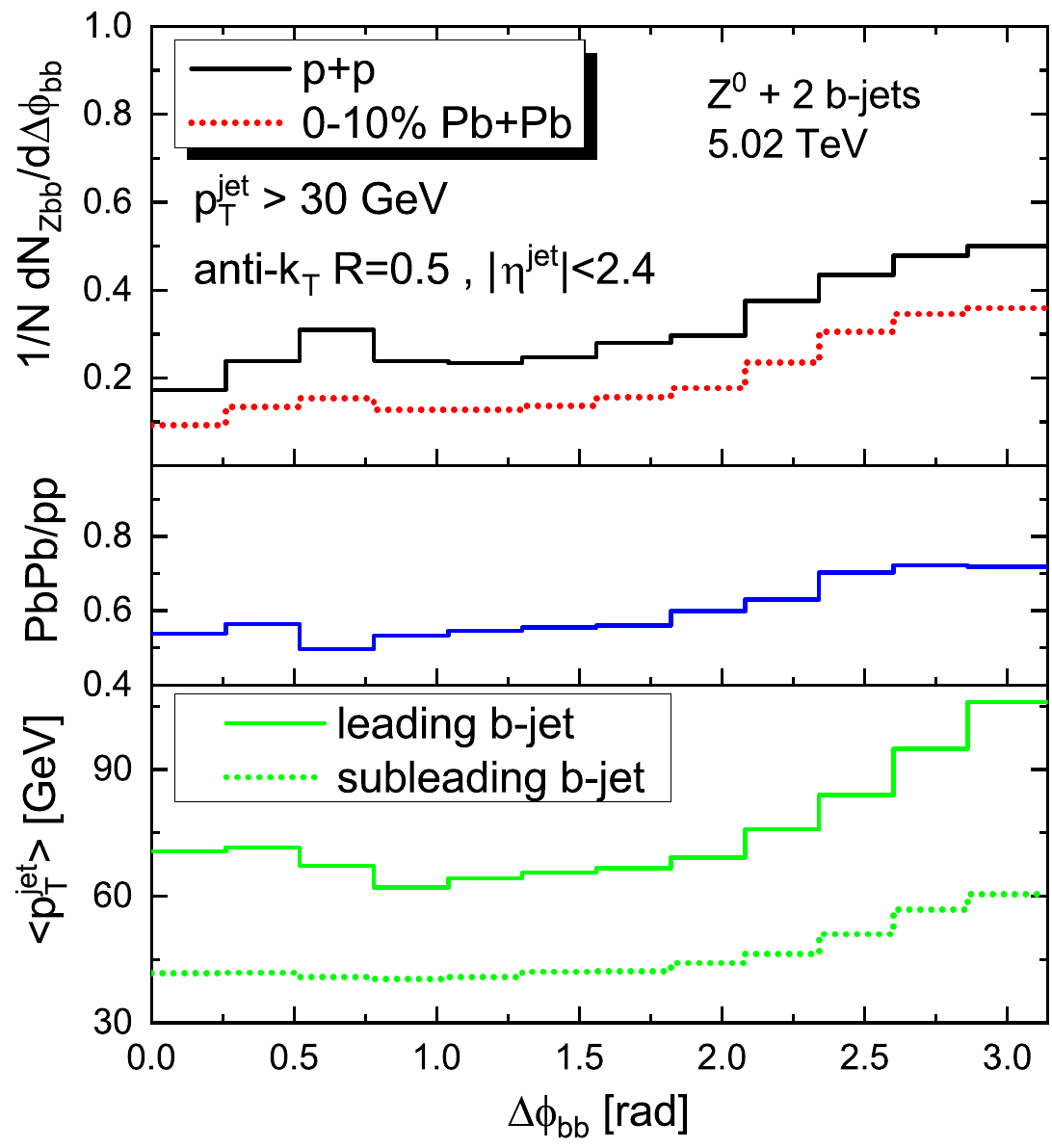}
\vspace*{0.2in}
\caption{Distributions of the azimuthal angular separation $\Delta\phi_{bb}$ of the two b-jets tagged by $Z^0$ boson both in p+p and $0-10\%$ Pb+Pb collisions at 5.02~TeV, the distributions are scaled by the $Z^0\,+\,2\,$b-jets event number $N$ in p+p collisions. Middle panel: ratio of the azimuthal angular separation in Pb+Pb to that in p+p. Bottom panel: averaged transverse momentum of the leading and the sub-leading b-jet as a function of $\Delta\phi_{bb}$.}
  \label{fig:phibbaa}
\end{center}
\end{figure}

As mentioned in Sec.~\ref{sec:ppbaseline}, the azimuthal angular separation $\Delta\phi_{bb}$ of the two b-jets tagged by $Z^0$ boson is also a useful observable to distinguish the contribution from subprocesses where $Z^0$ boson is emitted from one of the final state b quark or gluon splitting~($g\rightarrow b\bar{b}$)~\cite{Chatrchyan:2013zja}, as shown in diagrams (c) and (d) of Fig.~\ref{fig:process}. Note that these two categories of contributions corresponding to the cases that the two b-jets are almost back-to-back or collinear. What interests us is how the $\Delta\phi_{bb}$ distribution of these two categories of $Z^0\,+\,2\,$b-jet would be modified in the QGP. As shown in the top panel of Fig.~\ref{fig:phibbaa}, we plot the $\Delta\phi_{bb}$ distributions both in p+p and $0-10\%$ Pb+Pb collisions at $\sqrt{s_{NN}}$=5.02~TeV and also plot the ratio PbPb/pp in the middle panel. In the upper panel, we can find a kink at $\Delta\phi_{bb}\sim\pi/5$ and a peak at $\Delta\phi_{bb}\sim\pi$ in Fig.~\ref{fig:phibbaa}, which present the two contributions of $Z^0\,+\,2\,$b-jets production: the two b-jets are almost back-to-back or collinear. The two-peaks distribution is similar to what we have observed in the angular correlations of $b\bar{b}$ dijets~\cite{Dai:2018mhw}, and the peaks at smaller and larger $\Delta\phi_{bb}$ regions corresponding to the gluon splitting (GSP) processes and flavor creation (FCR) processes. In the middle panel, we observe an upward trend of the ratio from 0.5 to 0.7 as $\Delta\phi_{bb}$ increases. To figure out the $\Delta\phi_{bb}$ dependence of the ratio PbPb/pp, we estimate the initial jet  $\left\langle p_{T} \right\rangle$ of the leading and sub-leading b-jet in p+p collisions, as shown in the bottom panel of Fig.~\ref{fig:phibbaa}. We find that, for both the leading and subleading one, $\left\langle p_{T} \right\rangle$ is increasing with $\Delta\phi_{bb}$, which shows a similar trend with that of the ratio PbPb/pp versus $\Delta\phi_{bb}$. It may indicate that the medium modification of $\Delta\phi_{bb}$ in Pb+Pb has a close connection with the initial b-jet $\left\langle p_T \right\rangle$ distribution versus $\Delta\phi_{bb}$ in p+p collisions.

\begin{figure}[!t]
\begin{center}
\vspace*{-0.2in}
\hspace*{-.1in}
\includegraphics[width=3.2in,height=3.2in,angle=0]{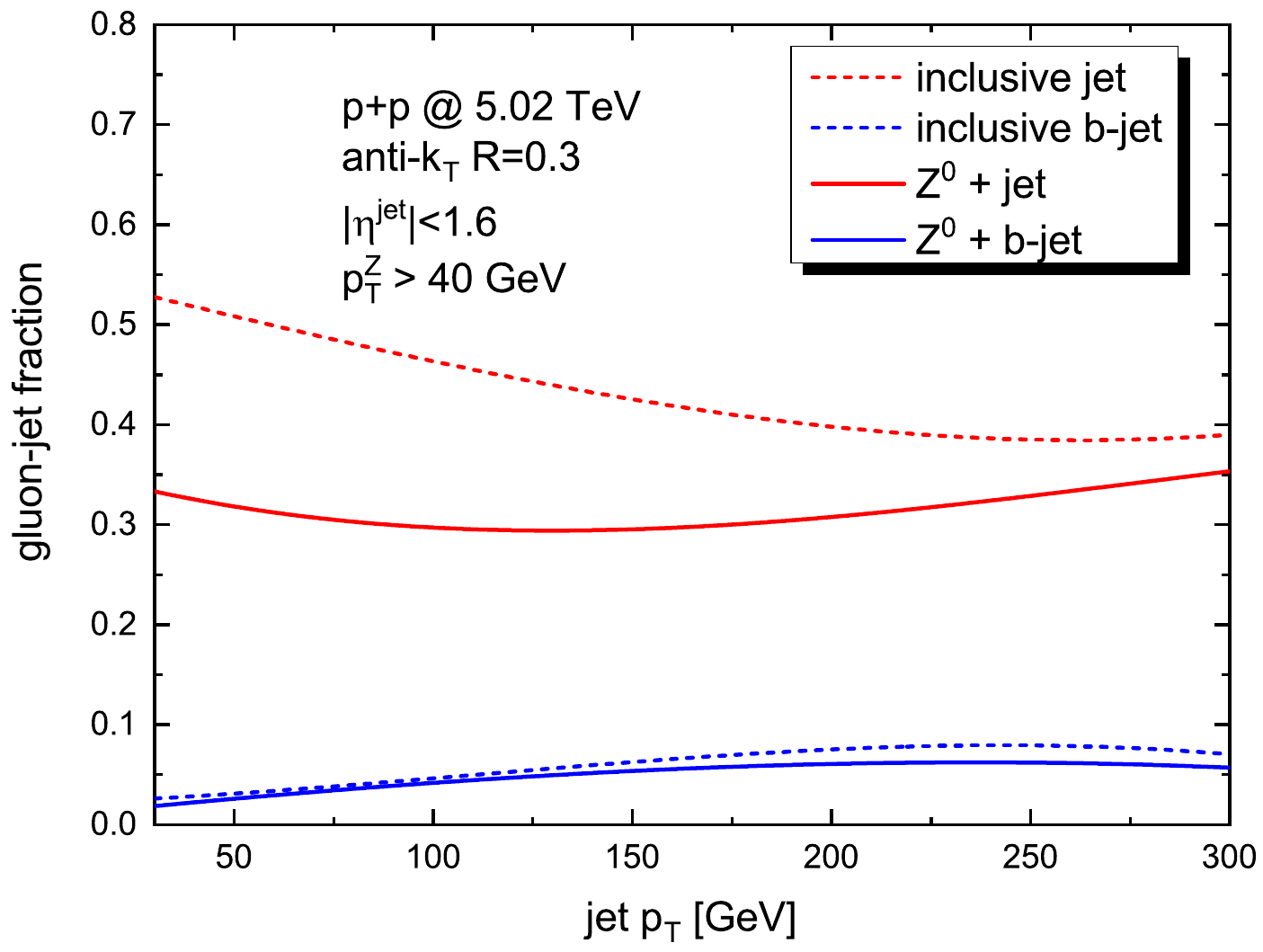}
\vspace*{0.2in}
\caption{Gluon initiated jet fraction as a function of transverse momentum of the inclusive jet~(red dash), the inclusive b-jet~(blue dash), and $Z^0\,+\,$jet~(red solid), $Z^0\,+\,$b-jet~(blue solid).}
\label{fig:frac}
\end{center}
\end{figure}

The associated production of $Z^0\,+\,$jet may shed new light on the mass dependence of the jet quenching effect in nuclear matter, owing to the high purity of light-quark-initiated jets. In Ref.~\cite{Kartvelishvili:1995fr}, the contributions from light-quark-jets and gluon-jets in the $Z^0\,+\,$jet production are approximately $70\%$ and $30\%$, respectively. To verify this point, in Fig.~\ref{fig:frac}, we estimate the gluon-jet fraction in four categories of jets in p+p collisions at 5.02~TeV: inclusive jet, inclusive b-jet, $Z^0$ tagged jet, and $Z^0$ tagged b-jet. The gluon-jet in the events can be identified by requiring that the gluon is the leading parton in the jets, which can be easily implemented in the FastJet program. We find that at $p_T^{\rm jet}\sim 50$~GeV, the gluon-jet fraction is approximately $50\%$ for the inclusive jet and $30\%$ for the $Z^0$ tagged jet. The $Z^0$-tagging requirement considerably decreases the gluon jet contribution (by $40\%$), especially at a low $p_T$.  More importantly, for the inclusive b-jet and $Z^0$ tagged b-jet, the contributions from gluon-jet are significantly suppressed owing to the requirement of b-quark tagging and show almost equal values.

\begin{figure}[!t]
\begin{center}
\vspace*{-0.2in}
\hspace*{-.1in}
\subfigure[]{\label{fig:xji}
  \epsfig{file=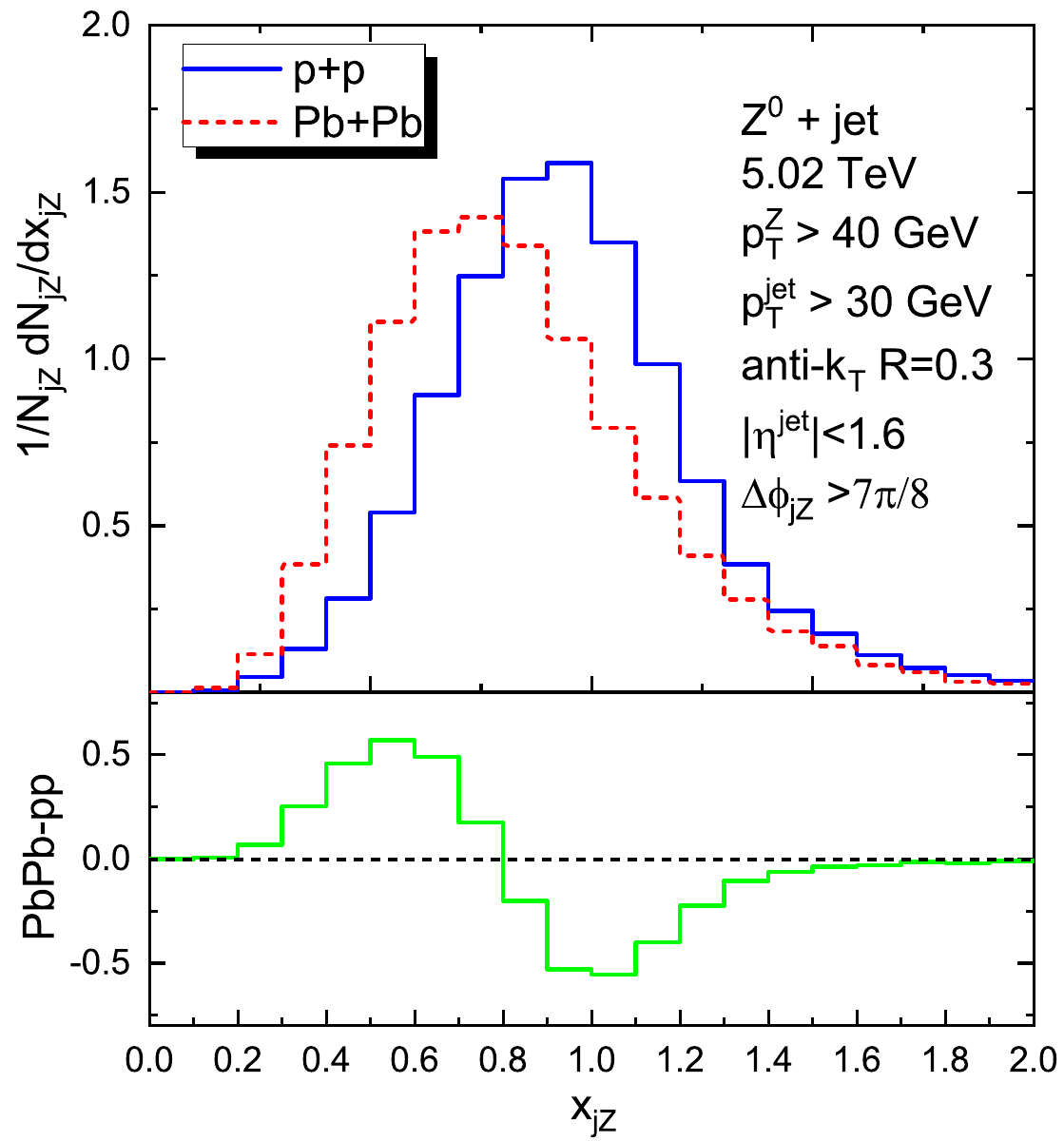, width=0.45\textwidth, clip=}}
  \subfigure[]{\label{fig:xjb}
  \epsfig{file=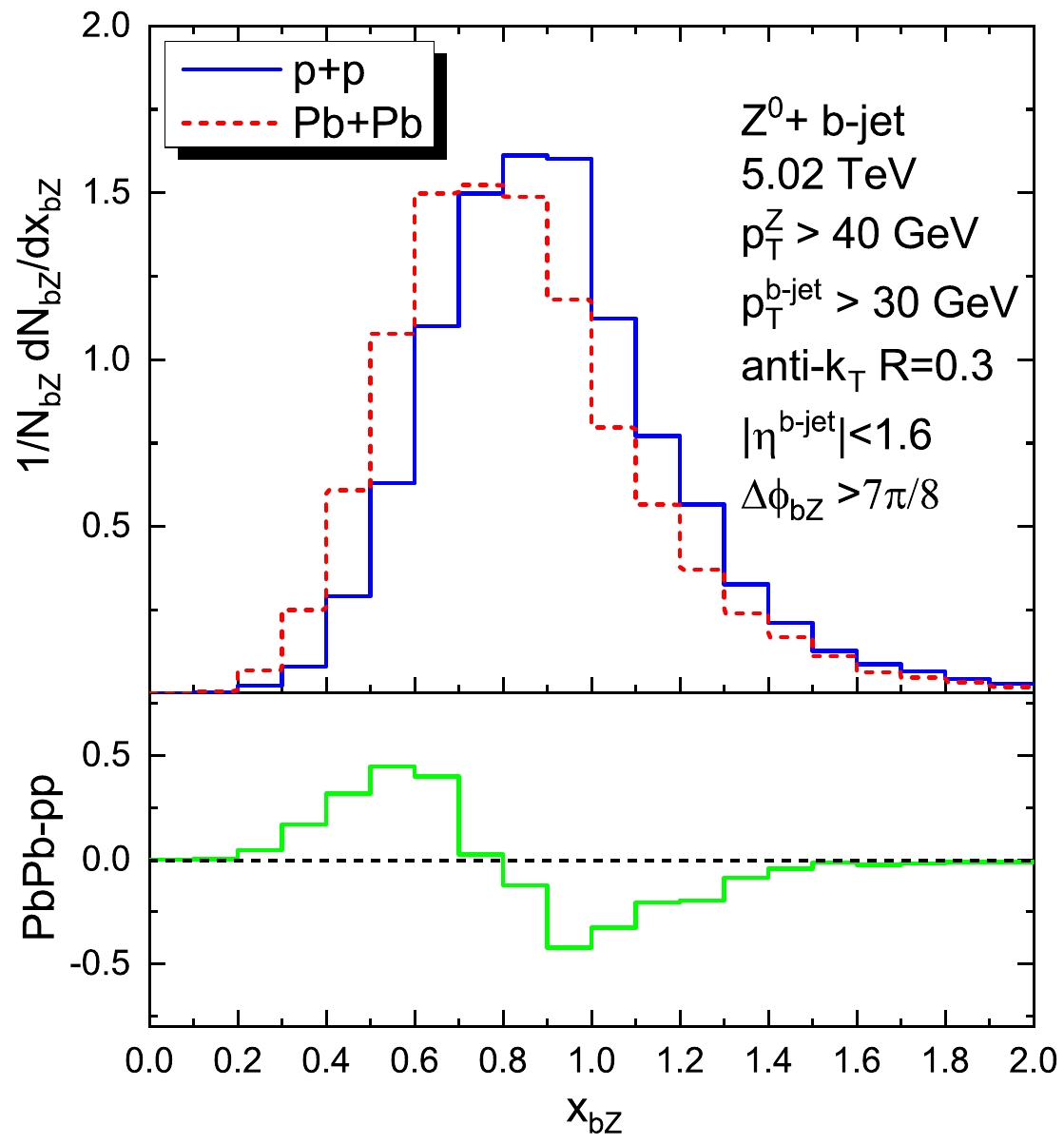, width=0.45\textwidth, clip=}}
\vspace*{0.1in}
\caption{(a) $x_{jZ}$ distributions of $Z^0\,+\,$jet in both p+p and $0-10\%$ Pb+Pb collisions at 5.02~TeV, as well as their difference in Pb+Pb to p+p. The distributions are normalized by the $Z^0\,+\,$jet number both in p+p and Pb+Pb collisions. (b) $x_{bZ}$ distributions of the $Z^0\,+\,$b-jet in both p+p and $0-10\%$ Pb+Pb collisions at 5.02~TeV, as well as their difference in Pb+Pb to p+p. The distributions are normalized by the $Z^0\,+\,$b-jet number in both p+p and Pb+Pb collisions.}
\end{center}
\end{figure}

\begin{table}
\begin{center}
\vspace*{0.2in}
\hspace*{0in}
\begin{tabular}{p{4.0cm}<{\centering}|p{2.0cm}<{\centering}|p{2.0cm}<{\centering}}
  \hline
  \hline
  \hspace*{0.1in}
  & $Z^0$+ jet & $Z^0+$ b-jet \\
  \hline
  ${\left\langle x_{J} \right\rangle }_{pp}$ & 0.987$\pm$0.0047 & 0.941$\pm$0.0056 \\
  \hline
  ${\left\langle x_{J} \right\rangle }_{PbPb}$ & 0.851$\pm$0.0061 & 0.849$\pm$0.0064 \\
  \hline
  $\Delta \left\langle x_{J} \right\rangle $ & 0.136$\pm$0.0108 & 0.092$\pm$0.012 \\
  \hline
  \hline
\end{tabular}
\caption{Mean value of momentum imbalance $x_J$ of $Z^0$ + jet and $Z^0$ + b-jet in both p+p  and $0-10\%$ Pb+Pb collisions at $\sqrt{s_{NN}}=5.02$~TeV, as well as the shifting of the mean value of momentum imbalance $\Delta x_{J}={\left\langle x_{J} \right\rangle }_{pp}-{\left\langle x_{J} \right\rangle }_{PbPb}$. The standard errors of $x_J$ in the simulations are also presented.}
\end{center}
\label{tab:axjz}
\end{table}

\begin{figure*}
\begin{center}
\vspace*{-0.2in}
\hspace*{-.1in}
\includegraphics[width=6.5in,height=2.5in,angle=0]{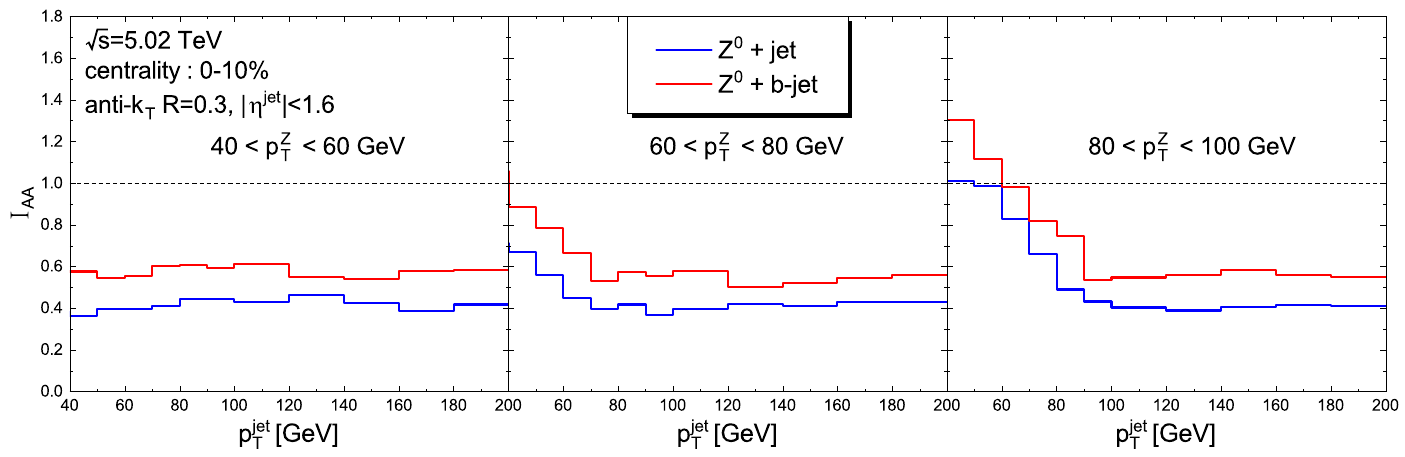}
\vspace*{0.1in}
\caption{Nuclear modification factor as a function of the transverse momentum of $Z^0$ + jet (blue line) and $Z^0$ + b-jet (red line) within three $p_T^Z$ ranges, i.e., $40-60$~GeV, $60-80$~GeV, and $80-120$~GeV, in $0-10\%$ centrality Pb+Pb collisions at $\sqrt{s_{NN}}=5.02$~TeV.}
\label{fig:IAA}
\end{center}
\end{figure*}

We plot the scaled $x_{jZ}$ distributions of $Z^0\,+\,$jet in both p+p and $0-10\%$ Pb+Pb collisions at $\sqrt{s_{NN}}$=5.02~TeV in Fig.~\ref{fig:xji} and with the same configuration, the $x_{bZ}$ distributions of $Z^0\,+\,$b-jet in Fig.~\ref{fig:xjb}. All the selected jets~(b-jets) must satisfy $\Delta\phi_{jZ}>7\pi/8$~($\Delta\phi_{bZ}>7\pi/8$) to guarantee that they are almost back-to-back with the $Z^0$ boson. We observe that the distributions of $x_J$ shift toward smaller values for both $Z^0\,+\,$jet and $Z^0\,+\,$b-jet in Pb+Pb collisions relative to their p+p baselines, owing to the energy loss of the tagged jets. To perform a more intuitive comparison between $Z^0\,+\,$jet and $Z^0\,+\,$b-jet, we show the difference of the $x_J$ distribution in Pb+Pb from that in p+p (see the bottom panels of Figs.~\ref{fig:xji} and Figs.~\ref{fig:xjb}), which have positive values at $0.2<x_{J}<0.8$ and negative values at $0.8<x_{J}<1.6$. Then we find that the absolute value of the difference~(PbPb-pp) of $Z^0\,+\,$jet is larger than that of $Z^0\,+\,$b-jet. Furthermore, we estimate the shifting of the mean value of momentum imbalance $\Delta {\left\langle x_{jZ} \right\rangle }
={\left\langle x_{jZ} \right\rangle }_{pp}-{\left\langle x_{jZ} \right\rangle }_{PbPb}$ for $Z^0\,+\,$jet, and $\Delta {\left\langle x_{bZ} \right\rangle }={\left\langle x_{bZ} \right\rangle }_{pp}-{\left\langle x_{bZ} \right\rangle }_{PbPb}$ for $Z^0\,+\,$b-jet, as shown in Tab.~I.

\begin{equation}
\left\langle x_{J} \right\rangle = \frac{1}{\sigma}\int\frac{d\sigma}{dx_{J}}x_{J}dx_{J}\, ,
\end{equation}

Here, $J$ denotes different processes. Note that the standard errors of $x_J$ in our simulations are also presented in Tab.~I. We find that within the statistical uncertainties, $\Delta\left\langle x_{jZ} \right\rangle$ ($\sim0.136\pm0.0108$) for $Z^0\,+\,$jet is visibly larger than $\Delta\left\langle x_{bZ} \right\rangle$ ($\sim0.092\pm0.012$) for $Z^0\,+\,$b-jet, indicating stronger modifications to the light-quark jet compared with the b-jet. Note that $x_{jZ}$~($x_{bZ}$) is defined by the ratio of jet~(b-jet) $p_T$ to $Z^0$ boson $p_T$. The shifting of $x_{jZ}$ ($x_{bZ}$) toward smaller values is directly related to the amount of jet~(b-jet) energy loss; hence, $\Delta\left\langle x_{jZ} \right\rangle>\Delta\left\langle x_{bZ} \right\rangle$ indicates that light quark jets lose more energy than b-jets.

The nuclear modification factor $I_{AA}$ is practically another good observable to address the mass hierarchy and flavor dependence of the jet quenching effect. Comparisons of $I_{AA}$ between $Z^0\,+\,$jet and $Z^0\,+\,$b-jet would provide more reliable evidence of the mass effect of jet quenching. For this purpose, we present the calculations of $I_{AA}$ of $Z^0\,+\,$jet and $Z^0\,+\,$b-jet in $0-10\%$ Pb+Pb collisions at $\sqrt{s_{NN}}$=5.02~TeV as a function of jet $p_T$ within three $p^Z_T$ windows in Fig.~\ref{fig:IAA}. First, we find different shapes of $I_{AA}$ in these three panels. In the left panel ($40<p_T^Z<60~$GeV) the two curves of $I_{AA}$ for $Z^0\,+\,$jet and $Z^0\,+\,$b-jet are flat, but in the right panel ($80<p_T^Z<120~$GeV) the curves show enhancement at $p_T^{\rm jet}<80$~GeV. This is because if we constrain $80<p_T^Z<120$~GeV in the event selection, the cross section at $p_T^{\rm jet}<80$~GeV is steeper, falling with jet $p_T$. Then, the jets shifting from higher $p_T$ to a lower $p_T$ because of the in-medium energy loss naturally lead to the relatively large $I_{AA}$ values in the lower $p_T^{jet}$ region, even larger than one. Additionally, we observe smaller values of $I_{AA}$ for Z + jet relative to those for Z + b-jet in the three panels of Fig.~\ref{fig:IAA}, which indicates that the yield of $Z^0$ tagged light-quark jets suffers stronger suppression after traversing the QCD matter than that of $Z^0$ tagged b-jets. Note that $I_{AA}$ is directly related to the $p_T$ shifting of the tagged jets due to in-medium energy loss. The smaller value of $I_{AA}$ for $Z^0$ + jet indicates a larger jet $p_T$ shift compared with that of $Z^0\,+\,$b-jet, suggesting that the energy loss of $Z^0$ tagged b-jets is smaller than that of $Z^0$ tagged light-quark jets. Our conclusions are consistent with the recent ATLAS measurements \cite{ATLAS:2022fgb}, which indicate that the $R_{AA}$ of the b-jet is higher than that of the inclusive jet. We hope that our complementary predictions can be tested by future measurements at the LHC, which may be helpful for solving the puzzle of the mass hierarchy of jet quenching.

\section{Summary}
\label{sec:summary}
Vector boson-tagged heavy quark jets are promising new tools for studying the jet quenching effect. In this work, we present a Monte Carlo transport simulation, which takes into account the elastic and inelastic jet interactions within a hydrodynamic background, to study the in-medium modification of $Z^0$ tagged b-jets. The NLO+PS event generator SHERPA was been used to provide the p+p baseline of $Z^0\,+\,$b-jet production, which agreed well with the CMS measurements. This framework has been proven to give good descriptions of medium modifications of $\Delta\phi_{jZ}$ and  $x_{jZ}$ of $Z^0\,+\,$jet, as well as the $R_{AA}$ of the inclusive b-jet, measured in Pb+Pb collisions at the LHC.

The angular correlation between the vector boson and heavy quark-tagged jets may be a new promising observable for studying the in-medium jet interaction. We present the first calculation of the azimuthal angular correlation $\Delta\phi_{bZ}$ of $Z^0\,+\,$b-jet in both p+p and $0-10\%$ Pb+Pb collisions at $\sqrt{s_{NN}}=$~5.02~TeV. We observe a flat suppression factor versus $\Delta\phi_{bZ}$, in contrast to the case of $Z^0\,+\,$jet, because the requirement of b-tagging excludes the contribution from multiple-jet processes. Additionally, we calculate the medium modification of the azimuthal angular correlation $\Delta\phi_{bb}$ in central Pb+Pb collisions at $\sqrt{s_{NN}}=5.02$~TeV and observe stronger suppression in a smaller $\Delta\phi_{bb}$ region of the distribution relative to that at $\Delta\phi_{bb}\sim\pi$. By analyzing the $\left\langle p_T \right\rangle$ distribution of the tagged b-jets, we find that the medium modification pattern on $\Delta\phi_{bZ}$ ($\Delta\phi_{bb}$) in Pb+Pb has a close connection with the initial $\left\langle p_T \right\rangle$ distribution versus $\Delta\phi_{bZ}$ ($\Delta\phi_{bb}$) in p+p collisions. These investigations may help us to understand the experimental measurements of jet angular correlations at the LHC in recent years, e.g., for the $\gamma+$jet and $Z^0+$jet.

Finally, we predict that the mass effect of jet quenching can be addressed by comparing the medium modifications of $Z^0\,+\,$jet and $Z^0\,+\,$b-jet. With the high purity of the quark jet in $Z^0\,+\,$(b-)jet events, we estimate the medium modification of the transverse momentum imbalance $x_{jZ}$ ($x_{bZ}$) and the nuclear modification factor $I_{AA}$ for both $Z^0\,+\,$jet and $Z^0\,+\,$b-jet in Pb+Pb collisions. We find a larger shift of $x_{jZ}$ and stronger suppression for $I_{AA}$ of $Z^0\,+\,$jet than for $Z^0\,+\,$b-jet indicating that b-jets lose less energy than light quark jets. These predictions can be tested via future measurements at the LHC and may provide a key to solving the puzzle of the mass hierarchy of jet quenching.

\acknowledgments
This research is supported by the Guangdong Major Project of Basic and Applied Basic Research No. 2020B0301030008, the Science and Technology Program of Guangzhou No. 2019050001, and the NSFC of China with Project Nos.~11935007, 12035007, 12247127. S. Wang is supported by China Postdoctoral Science Foundation under project No. 2021M701279.

\end{document}